\documentclass[12pt]{article}
\usepackage{amsmath,amsfonts,amssymb,amsxtra}

\usepackage{graphicx}
\usepackage[numbers,sort&compress]{natbib}
\usepackage{hypernat}
\usepackage{hyperref}
\numberwithin{equation}{section}

\newcommand{\preprint}[1]{\begin{table}[t]  %%
             \begin{flushright}               %%
             {#1}                             %%
             \end{flushright}                 %%
             \end{table}}                     %%
\renewcommand{\title}[1]{\vbox{\center\LARGE{#1}}\vspace{5mm}}
\renewcommand{\author}[1]{\vbox{\center#1}\vspace{5mm}}
\newcommand{\address}[1]{\vbox{\center\em#1}}
\newcommand{\email}[1]{\vbox{\center\tt#1}\vspace{5mm}}

\newcommand {\la} {\left \langle}
\newcommand {\ra} {\right \rangle}
\newcommand {\lb} {\left (}
\newcommand {\rb} {\right )}

\newcommand {\CalN} {\mathcal N}

\newcommand {\BR}   {\mathbb R}
\newcommand {\BZ}   {\mathbb Z}

\newcommand {\ve}  {\varepsilon}

\newcommand {\lam}  {\lambda}

\newcommand{\g}{\mathfrak{g}}
\newcommand{\h}{\mathfrak{h}}

\newcommand {\p} {\partial}

\DeclareMathOperator{\tr} {tr}

\DeclareMathOperator{\Pexp} {Pexp}

\newcommand{\SU}{SU}

\begin{document}

\unitlength = .8mm

\bibliographystyle{utphys}

\begin{titlepage}
\begin{center}
\hfill \\
\hfill \\

\preprint{pi-qf\&strings-277}
\preprint{ITEP-TH-24/12}

\title{
Correlators of Wilson loops and local operators\\from multi-matrix models and strings in AdS}

\renewcommand{\thefootnote}{\fnsymbol{footnote}}
\author{Simone Giombi$^{1,a}$ and
Vasily Pestun$^{2,b,\,}$\footnotemark}
\footnotetext{On leave of absence from ITEP, 117218, Moscow, Russia}

\address{${}^1$Perimeter Institute for Theoretical Physics, Waterloo, Ontario, N2L 2Y5, Canada}
\address{${}^2$School of Natural Sciences, Institute for Advanced Study, Princeton, NJ 08540 USA}

\email{$^a$sgiombi@pitp.ca,
$^b$pestun@ias.edu}

\end{center}

\abstract{
We study correlation functions of Wilson loops and local operators in a
subsector of ${\cal N}=4$ SYM which preserves two
supercharges. Localization arguments allow to map the problem to a
calculation in bosonic two-dimensional Yang-Mills theory. In turn, this
can be reduced to computing correlators in certain Gaussian multi-matrix
models. We focus on the correlation function of a Wilson loop and two local operators, and solve the corresponding three-matrix model exactly in the planar limit. We compare the strong coupling behavior to string theory in $AdS_5\times S^5$, finding precise agreement. We pay particular attention to the case in which the local operators have large R-charge $J\sim \sqrt{\lambda}$ at strong coupling.}

\vfill

\end{titlepage}

\eject

\tableofcontents

%%% Local Variables: 
%%% mode: latex
%%% TeX-master: "main"
%%% End:

\newcommand{\gym}{g_{4d}}

\section{Introduction}
Exact results in non-abelian gauge theories are rare and clearly of great importance. In supersymmetric gauge theories, the powerful technique of localization allows sometimes for such exact results for certain observables preserving some fermionic symmetries of the theory. For example the exact expressions conjectured in \cite{Erickson:2000af, Drukker:2000rr} for the circular 1/2-BPS Wilson loop in ${\cal N}=4$ SYM were proved  using localization in \cite{Pestun:2007rz}, as well as extendend to a large class of ${\cal N}=2$ theories. 

In ${\cal N}=4$ SYM, a wide generalization of the 1/2-BPS circle to
lower supersymmetric Wilson loops of arbitrary shapes was introduced in
\cite{Drukker:2007dw,Drukker:2007yx, Drukker:2007qr} and then classified
in \cite{Dymarsky:2009si}. An interesting subfamily of that construction consists of operators supported on any loop on a two-sphere $S^2$ embedded into the $\mathbb{R}^4$ spacetime. Generically, these Wilson loops are 1/8-BPS, and it was conjectured in \cite{Drukker:2007dw,Drukker:2007yx, Drukker:2007qr} that their quantum correlators are exactly captured by a purely perturbative calculation in bosonic 2d Yang-Mills. The conjecture was later strongly supported by the localization calculation in \cite{Pestun:2009nn}, where it was shown that the path-integral with insertions of those loop operators localizes on a 2d gauge theory closely related to the Hitchin/Higgs-Yang Mills system \cite{Moore:1997dj, Gerasimov:2006zt, Gerasimov:2007ap}, which can be seen to be perturbatively equivalent to ordinary bosonic 2d Yang-Mills.\footnote{In \cite{Pestun:2009nn} the one-loop determinant for fluctuations around the localization locus was not computed. The conjecture of  \cite{Drukker:2007dw,Drukker:2007yx, Drukker:2007qr} follows if one assumes that the determinant is just unity. Hence the conjecture for the 1/8-BPS loops is not yet proved at the same level of rigour as \cite{Pestun:2007rz}.} The calculation in the 2d theory can be then mapped to certain Gaussian multi-matrix models, which allow for an exact evaluation of the correlators. Several checks of the relation to 2d YM have appeared \cite{Young:2008ed, Bassetto:2008yf, Giombi:2009ms, Bassetto:2009rt, Bassetto:2009ms, Kalousios:2011hc}.\footnote{All calculations in the literature agree with the original conjecture except for a certain disagreement found in \cite{Bassetto:2009ms}. It would be good to clarify the nature of that disagreement.} In particular, recently the localization result for the expectation value of a Wilson loop in this family was used in \cite{Correa:2012at} (see also \cite{Fiol:2012sg}) to derive an exact expression for the low-angle limit of the cusp anomalous dimension. This has been checked using integrability up to three loops in \cite{Correa:2012hh} (see also \cite{Drukker:2012de}) and analytically to all loops in \cite{Gromov:2012eu}, providing a first link between localization and integrability results.  

The calculation in \cite{Pestun:2009nn} also suggested that localization applies in fact not only to the Wilson loops, but to a larger sector of operators that are annihilated by the same supercharge. This include certain chiral primary operators inserted on the $S^2$ \cite{Giombi:2009ds} as well as 't Hooft loops linked with the $S^2$ \cite{Giombi:2009ek}. The correlation function of a Wilson loop and a local operator in this sector was computed in \cite{Giombi:2009ds}, giving support and generalizing the original conjecture of \cite{Semenoff:2001xp} for the correlator of a 1/2-BPS Wilson loop and a chiral primary (see also \cite{Zarembo:2002ph} for the study of the large R-charge limit of this correlator). In \cite{Giombi:2009ek}\cite{Bassetto:2010yc}, the exact results implied by the relation to 2d were also used to obtain some explicit tests of the S-duality symmetry of the ${\cal N}=4$ SYM theory. 

In this paper, we continue the study of this supersymmetric subsector
and concentrate on  
mixed correlation functions of Wilson
loops and local operators. In particular, we focus on the limit in which
the local operators have large R-charges $J \sim \sqrt{\lambda}$ in the
strong coupling regime $\lambda\gg 1$. A motivation to look at this
problem is the recent progress in computing correlation functions of
operators with large charges using semiclassical strings in AdS (see
e.g. \cite{Janik:2010gc, Buchbinder:2010vw, Zarembo:2010rr,
  Costa:2010rz, Roiban:2010fe, Janik:2011bd, Kazama:2011cp,
  Buchbinder:2011jr, Escobedo:2010xs, Escobedo:2011xw, Gromov:2011jh}
and references therein). We hope that studying similar correlators in a
subsector where exact results are possible may provide detailed tests of
the gauge/string dictionary and serve as a useful benchmark for
various computations involving more general non-protected
operators.\footnote{Some examples of (non-supersymmetric)
   
correlation functions of Wilson loop and local operators in the
  limit of large charges were recently studied also in
  \cite{Alday:2011pf}\cite{Hernandez:2012zj}.} 

As mentioned above, the localization to 2d allows to map the calculation to certain multi-matrix models. Atthree-point level, we have to solve a Gaussian three-matrix model. Since we could not find explicit results available in the literature,\footnote{There is a vast literature on multi-matrix models, see e.g. \cite{Ginsparg:1993is}\cite{DiFrancesco:1993nw}\cite{eynard-1997}. The class of models that can in general be solved exactly are those of open chain type, where the interactions among the $n$ matrices are of the form $X_i X_{i+1}, i=1,...,n$. In our case we also encounter matrix models with closed chain interactions which are not solved in general. However, since our models are Gaussian, a solution is still possible, as we show.} here we present in detail a derivation of the exact planar resolvent for the most general Gaussian three-matrix model. The result can be applied to all possible three-point correlators of Wilson loops and local operators, but in this paper we focus our attention to the case of $\langle W O_{J_1}O_{J_2}\rangle$.  After deriving an exact prediction for this correlator, we study its strong coupling behavior in the regimes $J_1,J_2\ll 1$, $J_1\sim \sqrt{\lam}\gg J_2$ and $J_1,J_2\sim \sqrt{\lambda}$. In the first case, in particular, we observe a factorization at leading order at strong coupling which is completely analogous to the one seen in \cite{Buchbinder:2010ek} for four-point functions of two ``heavy'' and two ``light'' operators. For  $J_1,J_2\sim \sqrt{\lambda}$ we derive a prediction for the exponential behavior of the correlator which should be matched against a semiclassical string solution with two spikes which end on two boundary points. 

On the string theory side, we first derive a generalization of the solution of \cite{Zarembo:2002ph} describing $\langle W O_{J}\rangle$ for $J\sim \sqrt{\lambda}\gg 1$ to the case of the 1/4-BPS circular loop. The corresponding string lies in $AdS_3\times S^3$ and preserves less supersymmetries than the solution of \cite{Zarembo:2002ph}. Computing its area we find precise agreement with the two-matrix model derived from the 2d YM description. We then use this solution to obtain the string prediction for $\langle W O_{J_1}O_{J_2}\rangle$ at $J_1\sim \sqrt{\lambda}\gg J_2$, again observing perfect agreeement with the exact solution of the three-matrix model in the appropriate limit. In this paper we do not find the string solution which should describe $\langle WO_{J_1}O_{J_2}\rangle$ in the regime $J_1,J_2\sim \sqrt{\lambda}$. We derive however a set of first order differential equations which follow from supersymmetry of the system and which should be satisfied by all string solutions in $AdS_3\times S^3$ dual to the general $\langle WW\cdots OO\cdots\rangle$ correlators in our sector. Hopefully, these equations will be useful to find or characterize new explicit solutions.  

The paper is organized as follows. In Section 2, we review the Wilson loops and local operators of interest as well as the dictionary relating them to the 2d theory. In Section 3 we show how to derive from 2d YM the Gaussian multi-matrix models capturing mixed correlators of Wilson and local operators (in particular $\langle W OO\rangle$). In Section 4 we solve the general Gaussian three-matrix model in the planar limit, obtain an exact prediction for $\langle W OO\rangle$ and study its strong coupling limit. In Section 5 we derive the string solution dual to $\langle W_{1/4-BPS}O_{J}\rangle$ at $J\sim\sqrt{\lambda}\gg 1$, compare its area to the localization prediction and use it to compute $\langle W OO\rangle$ in the limit of one ``heavy'' and one ``light'' local operator. Finally, we study the supersymmetry constraints on the string solutions in our sector and derive a set of first order differential equations that the solutions should obey.

\section{Review of supersymmetric subsector and relation to 2d}
In our conventions, the $\CalN=4$ SYM action on $\BR^4$ with the standard flat metric is
\begin{equation}
  S_{SYM} = -\frac 1 {g_{YM}^2} \int d^4 x \lb \frac 1 2 \tr  F_{\mu \nu} F_{\mu \nu} + \tr D_{\mu} \Phi_A D_{\mu} \Phi_A + \dots \rb\,,
\end{equation}
where $\mu=1,\ldots,4$ are space-time indices and $A=1,\ldots, 6$ are $SO(6)_R$ indices. The covariant derivative is $D = d + A$, the curvature is
$F_{\mu\nu}=[D_{\mu}, D_{\nu}]$, all fields take value in the Lie
algebra of the gauge group $U(N)$ and represented by
anti-Hermitian matrices
 $A_{\mu} = A^a_{\mu} T_a,
\Phi=\Phi^a T_a$. The anti-Hermitian generators satisfy $\tr T_{a} T_{b} = -\frac 1 2 \delta_{a b}$, hence the action may be also written as 
\begin{equation}
  \label{eq:S4d}
  S_{SYM} = \frac 1 {2g_{YM}^2} \int d^4 x \lb \frac {1} {2} F^a_{\mu \nu} F^a_{\mu \nu} + D_{\mu} \Phi^a_A D_{\mu} \Phi^a_A + \dots \rb.
\end{equation}

The Wilson loops we study in this paper are the 1/8-BPS operators constructed in 
\cite{Drukker:2007qr,Drukker:2007yx,Drukker:2007dw}. They are supported on arbitrary closed curves on a $S^2$ inside $\mathbb{R}^4$ which we may define in Cartesian coordinates as
\begin{equation}
x_4=0\,,\qquad \sum_{i=1}^3 x_i^2=r^2\,.
\end{equation} 
The radius $r$ of the two-sphere is arbitrary, and we will set it henceforth to 1 for simplicity (the radius dependence is easily reintroduced if desired). The 1/8-BPS Wilson loops couple to three of the six scalars, $\Phi_i,\,i=1,2,3$, and  for any loop ${\cal C}\in S^2$, they are given by
\newcommand{\dimR}{d_R}
\newcommand{\dimRpr}{d_{R'}}
\begin{equation}
  \label{eq:Wilson-loop}
  W_{R} ({\cal C}) = \frac{1}{\dimR}\tr_{R} \Pexp \oint_{\cal C} (A_i  + i \ve_{ijk} \Phi_i x^k ) dx^j\,,
\end{equation}
where $\dimR$ denotes the dimension of the representation $R$. In other words, given a loop defined by $\vec{x}(s), \vec{x}^2=1$, the operator couples to the combination of scalars $(\vec{x}\times \vec{\dot{x}})\cdot \vec{\Phi}$. For arbitrary curve, these operators preserve four supercharges.\footnote{In flat space, these supercharges are linear combinations of Poincar\'e and superconformal supersymmetries.} Supersymmetry can be enhanced for special shapes. For example, the well-known 1/2-BPS circular Wilson loop is obtained by taking ${\cal C}$ to be an equator of $S^2$. Circles of arbitrary radius along latitudes of $S^2$ are 1/4-BPS and they coincide with the 1/4-BPS Wilson loops of \cite{Drukker:2006ga}. Note that since the four supercharges preserved by the loops do not depend on the contour, an arbitrary collection of Wilson loops on $S^2$ is also 1/8-BPS. 

As shown in \cite{Giombi:2009ds}, see also \cite{Pestun:2009nn}, it is possible to add an arbitrary number of local operators on the same $S^2$ while still preserving two supercharges. The relevant local operators are the following  
\begin{equation}
O_J(x)=\tr \left(x^i\Phi^i+i\Phi^4\right)^J\qquad x^i\in S^2\,, i=1,2,3
\label{eq:local}
\end{equation}
Note that these can be viewed as ordinary chiral primaries inserted at a specific point, where the orientation in the scalar space is correlated with the position of the operator. It is easy to see from the definition that the two-point function of these operators is position independent. In the planar limit we have
\begin{equation}
\langle O_{J}(x) O_{J'}(x')\rangle = J \left(\frac{\lambda}{16\pi^2}\right)^J \delta_{JJ'}\,.
\label{2ptOO}
\end{equation}
In fact, (\ref{eq:local}) are a special case of the superprotected operators introduced in \cite{Drukker:2009sf}, where it was shown that all the $n$-point functions $\langle O_{J_1}(x_1) O_{J_2}(x_2)\cdots O_{J_n}(x_n)\rangle$ are position independent and moreover tree-level exact. 

The system of any number of these local operators on $S^2$ preserves four supercharges \cite{Drukker:2009sf}\cite{Giombi:2009ds}. When the Wilson loops (\ref{eq:Wilson-loop}) are also present, the combined system is invariant under two supercharges \cite{Giombi:2009ds}.
While all $n$-point functions of the $O_J$'s are protected, mixed correlation functions of Wilson loops and local operators can have a non-trivial coupling dependence and will be the focus of this paper. 

A linear combination of the two supercharges preserved by the system is precisely the fermionic charge used in the localization calculation of \cite{Pestun:2009nn}, which therefore also applies to mixed correlators of Wilson loops and local operators. Hence, extending the conjecture of \cite{Drukker:2007qr,Drukker:2007yx,Drukker:2007dw}, it was proposed in \cite{Giombi:2009ds} that correlators of any number of Wilson loops and local operators can be computed exactly by the bosonic two-dimensional Yang-Mills theory on $S^2$ with action 
\begin{equation}
  \label{eq:S_{2d}}
  S_{YM_2} = \frac 1 {2g_{2d}^2} \int d^2 \sigma \sqrt{g}  \lb  \frac {1} {2} \tilde F^a_{\mu \nu} \tilde F_a^{\mu \nu} \rb\,,
\end{equation}
with the following 4d/2d map
\begin{equation}
W_R({\cal C}) \quad \leftrightarrow \quad \tr_R {\rm P} e^{\oint_{\cal C} \tilde A}\,,\quad \quad 
O_J(x) \quad \leftrightarrow \quad \tr \left(i*_{2d}\tilde F\right)^J\,,\quad \quad
g_{2d}^2 = - \frac {g_{4d}^2} {2\pi} 
\label{2d4d}
\end{equation} 
where we used tilde to denote the two-dimensional fields. Under this
map, on the 2d YM side we should extract only the perturbative, or zero-instanton, contribution.  In fact, non-trivial instantons on the 2d side also have an interpretation in 4d: they correspond to turning on 1/2-BPS 't Hooft
loop operators along a $S^1$ linked with the $S^2$ \cite{Giombi:2009ek}
(see also \cite{Bassetto:2010yc}). In this paper we will not consider 't
Hooft loops and focus on Wilson loops and local operators. One way to
extract perturbative contribution of the exact 2d YM quantities
\cite{Migdal:1975zf, Blau:1991mp, Blau:1993hj, Witten:1991we,
  Witten:1992xu} is to decompose them into a sum of instanton sectors
and then pick the term with zero instanton number \cite{Bassetto:1998sr,
  Bassetto:1999dg, Staudacher:1997kn}. Another way is to simply sum up the 2d perturbative expansion around the trivial vacuum $\tilde A=0$ in a gauge in which the theory becomes free. 
\begin{figure}
\begin{center}
\includegraphics[width=40mm]{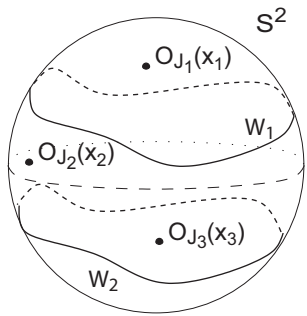}
\parbox{13cm}{\caption{Supersymmetric Wilson loops and local operators on $S^2$. The system of any number of loops of arbitrary shape and any number of local operators preserves 2 supercharges. Correlation functions of these operators are mapped by localization arguments to a multi-matrix model computation.}
\label{S2WsOs}}
\end{center}
\end{figure}
Either way, the end result is that the 4d correlators can be eventually mapped to certain Gaussian multi-matrix models
\begin{eqnarray}
&&\langle W_{R_1}({\cal C}_1)W_{R_2}({\cal C}_2)\cdots O_{J_1}(x_1)O_{J_2}(x_2)\cdots \rangle_{4d} \\
&&=\frac{1}{{\cal Z}}\int [dX_1][dX_2]\cdots [dY_1][dY_2]\cdots \,
\tr_{R_1}e^{X_1}\tr_{R_2}e^{X_2}\cdots \tr Y_1^{J_1}\tr Y_2^{J_2}\cdots\,
e^{-S_{m.m.}[X,Y]}\nonumber
\end{eqnarray}
where the matrix model action $S_{m.m.}[X,Y]$ is a quadratic form in $X_a,Y_a$ whose coefficients depend on the areas singled out by the Wilson loops and the topology of the system (i.e. in which regions the local operators sit, but not on their precise position). Note that we did not assume a large $N$ limit here. Localization, and hence the matrix model description, should apply at any finite $N$. 

As a special case, the Wilson loop expectation value is given by the 1-matrix model
\begin{equation}
\langle W_R({\cal C}) \rangle_{4d}=\frac{1}{\cal Z}\int [dX]\,\tr_R e^X\, e^{-\frac{A^2}{2g_{YM}^2A_1A_2}\tr X^2}
\label{W-vev}
\end{equation}
where $A_1,A_2$ are the areas singled out by the Wilson loop and $A=A_1+A_2=4\pi$. In particular for the 1/2-BPS circular loop $A_1=A_2=A/2$ and the correct Gaussian matrix model \cite{Erickson:2000af}\cite{Drukker:2000rr}\cite{Pestun:2007rz} is reproduced.  

Another example previously studied is the Wilson loop/local operator correlator, which using the map to 2d YM can be shown to be given by \cite{Giombi:2009ds}
\begin{equation}
\langle W_R({\cal C}) O_J(x)\rangle_{4d}=
\frac{1}{Z} \int [dX][dY] \, \tr_R e^X \,\tr Y^J\, e^{-\frac{A^2}{2 g^2_{YM}}
\tr \left(\frac{A_1}{A_2}Y^2-\frac{2i}{A_2}XY\right)}\,,
\label{WO-2mm}
\end{equation}  
where we have assumed that the local operator is in the region of area $A_1$. 
This result in particular reproduces and generalizes the conjecture of \cite{Semenoff:2001xp} for the exact correlator of the 1/2-BPS circular loop and a chiral primary (see also \cite{Semenoff:2006am} for the generalization to the 1/4-BPS circle).\footnote{Conformal symmetry fixes the correlation functions of a circular Wilson loop and a scalar primary operator up to an undetermined function of the coupling, see e.g. \cite{Gomis:2008qa, Alday:2011pf}. Hence the exact $\langle W O\rangle$ correlator on the sphere is enough to determine the correlator for arbitrary position of the local operator.}  The correlator of two Wilson loops on $S^2$ was also studied, see \cite{Giombi:2009ms}\cite{Bassetto:2009rt}. 

Note that the position independence and tree-level exactness \cite{Drukker:2009sf} of the correlation functions of local operators (\ref{eq:local}) can also be easily seen from the point of view of the 2d theory.\footnote{One may derive for example a Ward identity by acting with the differential on the correlation functions of $\tr (i*_{2d}\tilde F)^J$ operators  to show their position independence. Tree level exactness follows from the fact that 2d YM becomes Gaussian in an appropriate gauge.} The explicit multi-matrix model which computes the correlator in this case is given by \cite{Giombi:2009ds}
\begin{eqnarray}
\label{local-mm}
\langle O_{J_1} O_{J_2}\cdots O_{J_n}\rangle 
=\frac{1}{\cal Z}\int [dX_1 ][dX_2]\cdots[dX_n]&& e^{-\frac{8\pi^2}{g_{YM}^2}\tr\left(\frac{1}{n-1}(\sum_{a=1}^n X_a)^2-\sum_{a=1}^n X_a^2\right)}\times \cr 
&&\times \, \tr X_1^{J_1}\,  \tr X_2^{J_2}\, \cdots  \tr X_n^{J_n}\,.
\end{eqnarray}   
The matrix-model action is chosen to reproduce the tree-level propagator between the local operators, and is such that the propagators from an operator to itself are set to zero (the operators are understood to be normal-ordered). Sometimes in this paper we will also consider operators normalized in the same way as ordinary chiral primaries with unit two-point function, i.e.\footnote{A chiral primary ${\cal N}_J \tr (u\cdot \Phi)^J$ with $u^2=0$ and $u\cdot u^*=1$ has unit normalized 2-point function in the planar limit if ${\cal N}_J=2^{J/2}\left(-i\frac{2\pi}{\sqrt{\lambda}}\right)^J\frac{1}{\sqrt{J}}$. 
The factor of $i$ in the normalization factor is due to our convention that the gauge group generators are anti-hermitian.} 
\begin{equation}
\begin{aligned}
{\cal O}_J = \left(-i\frac{2\pi}{\sqrt{\lambda}}\right)^J\frac{1}{\sqrt{J}}\,\tr \left(x^i\Phi^i+i \Phi^4\right)^J\equiv {\cal N}_J O_J\,.
\label{norm-O}
\end{aligned}
\end{equation}

Let us mention that the localization arguments reviewed above should also apply when the local operators are inserted along the loop, i.e. for gauge invariant operators of the form 
\begin{equation}
\begin{aligned}
&\tr \left\{(x_1^i\Phi^i(x_1)+i \Phi^4(x_1))^{J_1}W[x_1,x_2](x_2^i\Phi^i(x_2)+i \Phi^4(x_2))^{J_2}W[x_2,x_1]\right\}\\  
&W[x_1,x_2]={\rm P} e^{\int_{x_1}^{x_2}(A_i  + i \ve_{ijk} \Phi_i x^k ) dx^j}
\end{aligned}
\end{equation}
and analogous operators with arbitrary number of insertions along the loop. These should map to the corresponding operators in 2d YM according to (\ref{2d4d}). We leave a detailed study of these operators for the future.

\section{Multi-matrix models from 2d YM} 
In this section we show how to derive from 2d YM the Gaussian
multi-matrix models computing the correlation functions of Wilson loops
and local operators. First, we will use perturbation theory in the
light-cone gauge to obtain the 3-matrix models for the explicit example
of a triple trace correlator of a 1/8 BPS Wilson loop and two local operators on $S^2$. Then, we will present a simple formula giving the multi-matrix model for the most general multi-point correlators, and show how to derive it from the known exact solution of 2d Yang-Mills \cite{Migdal:1975zf, Blau:1991mp, Blau:1993hj, Witten:1991we, Witten:1992xu}. 

\subsection{Light-cone perturbatione theory}
As done in \cite{Giombi:2009ds}, a simple way to derive the matrix model is to look at the perturbative Feynman diagram expansion in the $A_{\bar z}=0$
gauge, where $z,\bar z$ denote complex coordinates on $S^2$ with metric (the radius is set to 1) 
$$
ds^2=\frac{4 dzd\bar z}{\left(1+z\bar z\right)^2}\,.
$$
In the $A_{\bar z}=0$ gauge there are no interactions and the 2d YM action becomes simply\footnote{Recall that we use conventions in which the gauge field is anti-hermitian, as in \cite{Giombi:2009ds}. Also, in this section we will omit tilde's on the 2d fields, since confusion with 4d fields will not arise.}
\begin{equation}
\label{eq:2daction-light-cone}
S_{YM_2}=\frac{1}{g_{2d}^2}\int d^2z \sqrt{g} g^{z\bar z} g^{z\bar z}\tr\left(\partial_{\bar z}A_z\partial_{\bar z} A_z\right)\,.
\end{equation}
We use notations $d^2 z = d \bar z \wedge dz = 2 i dx \wedge dy$ for $z = x + i y$, and $\sqrt{g} = -i g_{\bar z z}$,
so that $d^2 z \sqrt{g}$ is the conventional volume form on $S^2$ normalized as 
\begin{equation}
  \label{eq:normalization}
  \int {d^2 z \sqrt{g}} = 4 \pi.
\end{equation}
The gauge field propagator is
\begin{equation}
\la (A_z)^i_j(z) (A_z)^k_l (w)\ra = -\frac{g^2_{2d}}{2\pi} \delta^i_l \delta^k_j \frac{1}{1+z\bar z}\frac{1}{1+w \bar w}\frac{\bar z-\bar w}{z-w}\,.
\label{light-prop}
\end{equation} 
Using this, one gets the propagator for the field strength $i*_{2d}F=-\frac{1}{2} (1+z\bar z)^2\partial_{\bar z}A_z$
\begin{equation}
\langle i*_{2d}  F^i_j(z)\, i *_{2d}  F^k_l(w)\rangle
=-\delta^i_l \delta^k_j \left( \frac{g_{2d}^2 }{8\pi }-
\frac{i g^2_{2d} }{4}(1+z \bar z)^2 \delta^2(z-w)\right)\,.
\label{prop-2d}
\end{equation}
For convenience, let us record here also the propagator between the field strength and the gauge field 
\begin{equation}
\langle i*_{2d}  F^i_j(z)\,(A_z)^k_l(w)\rangle
=- \frac{g_{2d}^2}{4\pi}\delta^i_l \delta^k_j \frac{1}{1+w \bar w}\frac{1+z\bar w}{z-w}\,.
\label{prop-FA}
\end{equation}

\begin{figure}
\begin{center}
\includegraphics[width=80mm]{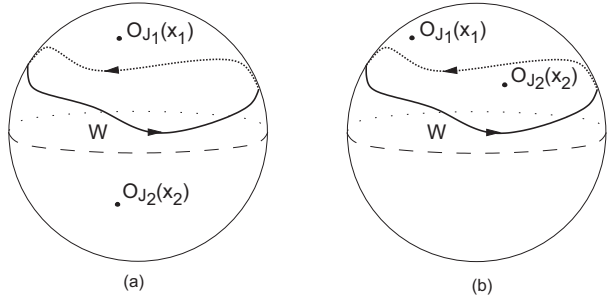}
\parbox{13cm}{\caption{The two distinct topologies for the correlator $\langle W OO\rangle$. Fig. (a) is mapped to an open chain 3-matrix model, while (b) to a closed chain one.}
\label{WOO-S2}}
\end{center}
\end{figure}

There are two distinct topologies for the correlator of a Wilson loop and two local operators on $S^2$: one where the operators are on opposite hemispheres compared to the loop and the other where the operators are on the same hemisphere, see Figure \ref{WOO-S2}.

Let us start with the case in which the two local operators are on
opposite hemispheres compared to the Wilson loop.
Using the area preserving invariance of 2d YM and the position independence, we can always choose the loop to
be a circle at some latitude angle, and place one local operator, say
$O_{J_1}$, at the north pole and the other, $O_{J_2}$, at the south
pole. Let us parameterize the loop as $z(\tau)= r_0 e^{i \tau}$,
where $r_0 =\tan \frac{\theta}{2}$ and $\theta_0$ is  the latitude
angle. It is then easy to see that the propagators in the $A_{\bar z}=0$ gauge are all constants
($\dot z \equiv \p_{\tau} z$):
\begin{equation}
\begin{aligned}
\label{eq:2dym-prop}
&\la \dot z_1 (A_z)^i_j(z_1) \dot z_2 (A_z)^k_l(z_2)  \ra =  -\delta^i_l \delta^k_j \frac{g_{2d}^2}{2\pi}\frac{r_0^2}{(1+r_0^2)^2}=-\delta^i_l \delta^k_j \frac{g^2_{2d}}{8\pi^2}\frac{A_1 A_2}{A}\,\\
&\la i *_{2d} F^i_j(0)\, \dot z (A_z)^k_l(z) \ra = -i\delta^i_l \delta^k_j  \frac{g_{2d}^2r}{4\pi}\frac{1}{1+r_0^2}= -i\delta^i_l \delta^k_j  \frac{g_{2d}^2}{4\pi}\frac{A_2}{A}\,\\
&\la i *_{2d} F^i_j(\infty)\, \dot z (A_z)^k_l(z) \ra = i\delta^i_l \delta^k_j  \frac{g_{2d}^2r}{4\pi}\frac{r_0^2}{1+r_0^2}= i\delta^i_l \delta^k_j  \frac{g_{2d}^2}{4\pi}\frac{A_1}{A}\,\\
&\la i *_{2d} F^i_j(\infty)\,i*_{2d} F^k_l(0)\ra = -\delta^i_l \delta^k_j  \frac{g_{2d}^2}{8\pi}\,.
\end{aligned}
\end{equation}
Since the propagators are all constants, the sum of Feynman diagrams is
obviously given by a matrix model. If we assign matrices
$X_1,X_2,X_3$ to respectively $O_{J_1},W,O_{J_2}$, the
non-vanishing matrix propagators are then,
using the relation $g_{2d}^2=-2/A g_{YM}^2=-\frac{1}{2\pi}g_{YM}^2$
\begin{equation}
\begin{aligned}
%\label{eq:MM-prop}
&\la (X_1)^i_j (X_2)^k_l \ra = i g_{YM}^2 \frac{ A_2}{A^2} \delta^i_l \delta^k_j \equiv \frac{1}{N}\lam_{12}\delta^i_l \delta^k_j\,,\\
&\la (X_2)^i_j (X_2)^k_l \ra = g_{YM}^2 \frac{A_1 A_2}{A^2} \delta^i_l \delta^k_j \equiv \frac{1}{N}\lam_{22}\delta^i_l \delta^k_j\,,\\
&\la (X_2)^i_j (X_3)^k_l \ra = -i g_{YM}^2 \frac{A_1}{A^2}\delta^i_l \delta^k_j \equiv \frac{1}{N}\lam_{23}\delta^i_l \delta^k_j\,,\\
&\la (X_1)^i_j (X_3)^k_l \ra = g_{YM}^2 \frac{1}{A^2}\delta^i_l \delta^k_j \equiv \frac{1}{N}\lam_{13}\delta^i_l \delta^k_j\,,
\label{open-chain-propagators}
\end{aligned}
\end{equation}
where we have introduced the shorthand notation $\lam_{ab}$ to denote the propagator from $a$ to $b$ matrix, $a,b=1,2,3$. Inverting the propagator, one gets the matrix model action
\begin{equation}
\label{eq:OWOaction}
S_{OWO}=\frac{A^2}{2g_{YM}^2}\tr\left(\frac{A_1}{A_2}X_1^2-\frac{1}{A_1A_2}X_2^2+\frac{A_2}{A_1}X_3^2
-\frac{2i}{A_2 }X_1 X_2 +\frac{2i}{A_1 }X_2 X_3\right)
\end{equation}
and the correlator is given by
\begin{equation}
\langle O_{J_1}W_R({\cal C}) O_{J_2}\rangle =\frac{1}{\cal Z}\int [dX_1 ][dX_2][dX_3]\, \tr X_1^{J_1}\, \tr_R e^{X_2}
\tr\, X_3^{J_2}\, e^{-S_{OWO}}\,.
\end{equation}
We see that only adjacent matrices interact. This is known in the literature as an open chain multi-matrix model. The same structure arises for the correlator of three Wilson loops with the topology of three latitudes on $S^2$, see \cite{Giombi:2009ms}\cite{Bassetto:2009rt}. In that case, the explicit three-matrix model is given by
\begin{eqnarray}
%\begin{aligned}
&&\langle W_{R_1}({\cal C}_1) W_{R_2}({\cal C}_2) W_{R_3}({\cal C}_3)\rangle =\frac{1}{\cal Z}\int [dX_1 ][dX_2][dX_3]
\, \tr_{R_1} e^{X_1}\, \tr_{R_2} e^{X_2}\,  {\rm tr}_{R_3} e^{X_3}
\, e^{-S_{WWW}}\,,\cr
&&S_{WWW}=\frac{A}{2g_{YM}^2}\tr \left(\frac{1}{A_1}X_1^2+\frac{1}{A_{12}}(X_1-X_2)^2+\frac{1}{A_{23}}(X_2-X_3)^2
+\frac{1}{A_3}X_3^2\right)\,.
\label{S-WWW}
%\end{aligned}
\end{eqnarray}
where $A_1,A_{12},A_{23},A_3$ are the areas of the regions singled out by the three Wilson loops. The generalization to the case of any number of Wilson loops with the topology of latitudes on $S^2$ is straightforward \cite{Giombi:2009ms}\cite{Bassetto:2009rt} and gives an open chain multi-matrix model (see also next section for the derivation of the most general correlator).

When the local operators are on the same hemisphere, a similar derivation goes through. 
Assuming that both local operators described by $X_1$ and $X_3$
are to the north of the Wilson loop described by $X_2$,
the  only  change compared to (\ref{open-chain-propagators}) is that now
\begin{equation}
\begin{aligned}
&\la (X_2)^i_j (X_3)^k_l \ra = i g_{YM}^2 \frac{A_2}{A^2} \delta^i_l \delta^k_j\equiv \frac{1}{N}\lam_{23} \delta^i_l \delta^k_j\\
&\la (X_1)^i_j (X_2)^k_l \ra=i g_{YM}^2 \frac{A_2}{A^2}\delta^i_l \delta^k_j\equiv \frac{1}{N}\lam_{12}\delta^i_l \delta^k_j .
\label{closed-chain-propagators}
\end{aligned}
\end{equation}
Then the matrix model action is
\begin{equation}
  \label{eq:S}
  S_{OOW} =
\frac{A^2}{2 g_{YM}^2 (A_1+2 A_2)}
\tr \left (
-A_2 X_1^2+\frac{X
   _2^2}{A_2}-A_2X_3^2-2 i X_1 X_2-2
   i X_2 X_3+2A X_3 X_1
\right )\,.
\end{equation}
In this case all pair of matrices interact, and we
may refer to this model as a closed chain.

\subsection{General multi-matrix model from 2d YM exact solution}
\label{constant-F}
In \cite{Giombi:2009ms}, a simple way to derive the multi-matrix model giving the correlator of several Wilson loops was proposed. The argument was based on the observation that \cite{Blau:1991mp,Witten:1992xu,Blau:1993hj} 2d YM on a Riemann surface $\Sigma$ localizes on classical configurations solving $d_{A} *F = 0$. Also, two dimensional Yang-Mills theory essentially reduces to the abelian theory~\cite{Blau:1991mp,Blau:1993hj}, so that one can take $*F$ to be a piecewise constant function on $\Sigma$ with jumps supported on the Wilson loops. The value of $*F$ in each region is then related via Stokes' theorem to the integral of the gauge field $\oint_{{\cal C}_i} A=iX_i$ along each loop, where the matrices $X_i$ are the variables in the multi-matrix model. Let us review how this works in the simplest case of a single Wilson loop expectation value on $S^2$. We divide the sphere in two regions $\Sigma_1$, $\Sigma_2$ of areas $A_1$, $A_2$. The field strength is taken to be 
\begin{equation}
*F=i\phi_1 \quad {\rm in} ~~\Sigma_1\,,\qquad *F=i\phi_2 \quad {\rm in}~~\Sigma_2\,,
\end{equation} 
where $\phi_1$, $\phi_2$ are constant hermitean matrices (we assume $U(N)$ gauge group here). Stokes' theorem gives
\begin{equation}
iX\equiv\oint_{\cal C} A = \int_{\Sigma_1}F=i A_1 \phi_1=-\int_{\Sigma_2}F=-i A_2 \phi_2\,.
\end{equation}
The 2d Yang-Mills action localized to constant curvatures is then
\begin{equation}
S=-\frac{1}{g^2_{2d}}\int d^2\sigma\sqrt{g}\tr (*F)^2=
\frac{1}{g^2_{2d}}\tr (\frac{1}{A_1}X^2+\frac{1}{A_2}(-X)^2)
\end{equation}
and we get the matrix model
\begin{equation}
\langle W \rangle = \frac{1}{Z}\int [dX]\tr_R e^{i X} e^{-\frac{A}{g^2_{2d}A_1A_2}\tr X^2}=\frac{1}{Z}\int [dX]\tr_R e^{X} e^{-\frac{A^2}{2g^2_{YM}A_1A_2}\tr X^2},
\end{equation}
where in the last step we have used the relation (\ref{2d4d}) between 2d
and 4d coupling constants and formally changed variables from $X$ to
$iX$. This is indeed the expected form of the familiar one-matrix model
(\ref{W-vev}). In this derivation one assumes that after abelianization
the gauge bundle is trivial, which precisely corresponds to dropping the
contribution of the unstable instantons \cite{Blau:1991mp,
  Blau:1993hj}. In other words, the matrix model derived in this way
 computes the perturbative contribution of the 2d YM observables 
as required by the conjecture of \cite{Drukker:2007yx, Drukker:2007qr, Pestun:2009nn}. 

The argument readily generalizes to the case of $k$ non
self-interesecting Wilson loops of arbitrary topology a
general Riemann surface $\Sigma$. We can always dissect $\Sigma$ 
 into several pieces $\{\Sigma_m\}$ which
have the topology of a sphere with one, two or three holes, such 
that all given non-intersecting Wilson loops run along some of the cuts. 
The localization to constant curvature gives the multi-matrix model action
\begin{equation}
S_{\rm m.m.}=\frac{1}{g^2_{2d}}\sum_{\{\Sigma_m\}} \frac{1}{A_{\Sigma_m}} \tr \left(\sum_{i\in \partial \Sigma_m}s_i^{(m)} X_i\right)^2,
\label{multi-m.m.}
\end{equation}
where $s_i^{(m)}=\pm 1$ is fixed by the orienation of the Wilson loops relative to $\Sigma_m$. The Wilson loop correlator is then
\begin{equation}
\langle W_{R_1}({\cal C}_1)\cdots W_{R_k}({\cal C}_k)\rangle =\frac{1}{Z}\int [dX_1]\cdots[dX_k]\,
\tr_{R_1}e^{i X_1}\cdots \tr_{R_k}e^{i X_k}\, e^{-S_{\rm m.m.}}\,.
\end{equation}
To compute mixed correlators of Wilson and local operators, we proceed as follows. We shrink the Wilson loop ${\cal C}_i$ which we want to replace by a local operator, and substitute in the above action  $iX_i  = a_i (*F_i)$, where $a_i$ is the small area of the corresponding disk. We then compute the propagators by inverting the kinetic operator, and take the limit of $a_i\rightarrow 0$ while killing the propagator from  $*F_i$ to
$*F_i$ itself to implement normal ordering. Inverting again the propagator matrix we then obtain the effective matrix model action for mixed correlators with normal ordering prescription in effect on local operators. It is easy to verify that this procedure agrees with the matrix models shown earlier for the special cases $\langle W O\rangle$, $\langle W OO\rangle$ which were derived by summing up light-cone perturbation theory.

While the localization to constant $F$ argument is rather intuitive and
convenient, it is not completely rigorous. For example, one may worry
about subtleties in applying Stokes' theorem in the non-abelian
case. For this reason, we explicitly show here that the matrix model
action (\ref{multi-m.m.}) indeed corresponds to the perturbative, or
zero-instanton sector, contribution to the exact 2d YM partition function \cite{Migdal:1975zf, Blau:1991mp, Blau:1993hj, Witten:1991we, Witten:1992xu}.  

To be completely general, let us consider 2d YM with arbitrary compact gauge group $G$ on a Riemann surface $\Sigma$ (the specialization to $G=U(N)$ is straightforward). In the temporal gauge $A_0=0$  the Lagrangian is $
\frac{1}{g^2} \dot A_1^2$, then the Hamiltonian is $\frac{1}{4} g^2
\Delta$ where $\Delta$ is the Laplacian equal to the second Casimir. The
exact partition function with possibly several Wilson loops inserted is
obtained by gluing the spheres with one, two or three holes along the boundaries \cite{Witten:1991we,Witten:1992xu})
\begin{equation}
\label{eq:Migdal}
\begin{aligned}
  K_{\rm 1}(U_1) = \sum_{\lambda} d_\lambda e^{-\frac{1}{4} g^2 A
    C_2(\lambda)} \chi_\lambda (U_1)\\
  K_{\rm 2}(U_1,U_2) = \sum_{\lambda} e^{-\frac 1 4 g^2 A C_2 (\lambda) }
  \chi_{\lambda}(U_1) \chi_{\lambda}(U_2)\\
  K_{\rm 3}(U_1,U_2,U_3) = \sum_{\lambda} \frac{1}{d_\lambda} 
  e^{-\frac 1 4 g^2 A C_2(\lambda)}\chi_\lambda(U_1) \chi_\lambda (U_2)  \chi_\lambda(U_3) 
\end{aligned}
\end{equation}
where $U_i$ are the holonomies of the connection on the boundary,
$\lambda$ labels irreducible representations of $G$ of dimension
$d_{\lambda}$, and $\chi_{\lambda}$ is the corresponding
character. $C_2(\lambda)$ denotes the value of the second casimir of the
Lie algebra of $G$ in the irrep $\lambda$, and $A$ is the 2d area of a
given piece. For example, consider a sphere with three Wilson loops of
not concentric topology. We can glue this configuration from a sphere
with three holes $K_3$ and three disks $K_{1}$. The Wilson loop correlator is obtained by including the insertions of $\chi_{R_i}(U_i)$ for each Wilson loop and integrating over the boundary holonomies. Explicit examples for gauge group $U(1)$ and $U(2)$ are given in appendix.

We now want to show that the matrix model with action (\ref{multi-m.m.}) is the perturbative contribution to the exact partition function obtained from the blocks (\ref{eq:Migdal}). For example, consider the block $K_3$ (the other
cases $K_2 , K_1$ are treated similarly).
Let the holonomies $U_i$ be parametrized by elements $x_i$ in the Cartan of
the Lie algebra of $G$, such that $U_i$ is a conjugate of an element
$e^{x_i}$ of the maximal torus of $G$. The character $\chi_\lambda$ of the
irrep with highest weight $\lambda$ is given by Weyl's formula 
\begin{equation}
  \chi_\lambda(x) = \frac{ \sum_{w \in {\cal W}} (-1)^{w} e^{   (\lambda+\rho,
      w(x))}}{R(x)}
\end{equation}
where ${\cal W}$ is the Weyl group, $R(x)$ is the Weyl denominator 
\begin{equation}
  R(x) = \prod_{\alpha > 0} (e^{\frac 1 2 (\alpha,x)} - e^{-\frac 1
    2 (\alpha,x)}),
\end{equation}
and $\rho$ is the Weyl special vector defined by the condition that
 $(\rho, \alpha_i) = 1$ for all simple roots.\footnote{
In finite-dimensional Lie algebras $\rho = \frac 1 2 \sum_{\alpha > 0} \alpha$.}
The dimension of the irrep with highest weight $\lambda$ is 
\begin{equation}
  d_{\lambda} =\prod_{\alpha > 0} 
 \frac{ (\alpha, \lambda + \rho)}{ (\alpha, \rho)}\,,
\end{equation}
where $\alpha > 0 $ denotes positive roots. 
The second Casimir $C_2(\lambda)$ is 
\begin{equation}
  C_2(\lambda) = (\lambda+\rho)^2 - \rho^2 \,.
\end{equation}
Let $\Delta(x)$ be the Weyl measure on the Cartan of the Lie
algebra of $G$  
\begin{equation}
  \Delta(x)  = \prod_{\alpha > 0} (\alpha, x)\,.
\end{equation}
Hence, up to an irrelevant constant factor, the building block $K_3$ is given by 
\begin{equation}
\label{eq:building-abel}
  K_3(x_1,x_2,x_3) = \sum_{\lambda \in P^{+}} 
\left(\frac{ 1} {\Delta(\lambda + \rho)} 
\prod_{i=1}^{3} \frac{    \sum_{w_i \in W}
 (-1)^{w_i} e^{ (\lambda+\rho, w_i(x_i))}}{R(x_i)} \right) 
 e^{ - \frac{g^2 A_{\rm tri}}{4} ( \lambda + \rho)^2}
\end{equation}
where $P^{+}$ denotes the subset of dominant weights in the weight
lattice of $G$.

The exact 2d YM partition function is the integral over the holonomies
associated with the loops that join the building blocks. Each loop
variable enters the measure of the integration with the two building
blocks associated with it
\begin{equation}
  Z = \int [dU_1] K_{\Sigma_m}(U_1, \dots, ) K_{\Sigma_n}(U_1, \dots, ) \dots 
\end{equation}
where $\Sigma_m$, $\Sigma_n$ are two building blocks (either disk, cylinder or triunion) which share a boundary loop. 
We can integrate over adjoint orbits of $G$ and reduce the integral 
to the maximal torus of $G$ on which the measure
of integration is
\begin{equation}
\label{eq:abel-measure}
  [dU_i] =  dx_i R(x_i)^2\,.
\end{equation}
The factors $R(x_i)^2$ in the measure cancel the denominators $R(x_i)$
in the blocks (\ref{eq:building-abel}). Let $d x_i$ be the flat abelian
measure of integration and $K^{\rm flat}$ be the blocks
(\ref{eq:building-abel}) without denominators, i.e. 
\begin{equation}
\label{eq:building-abel-flat}
  K_3^{\rm flat} (x_1,x_2,x_3) = \sum_{\lambda \in P^{+}} \left (
\frac{ 1} {\Delta(\lambda + \rho)} 
\prod_{i=1}^{3}  \left(  \sum_{w_i \in W}
 (-1)^{w_i} e^{ (\lambda+\rho, w_i(x_i))} \right) 
\right) e^{ - \frac{g^2 A_{\rm tri}}{4} ( \lambda + \rho)^2}
\end{equation}
so that the partition function is 
\begin{equation}
\label{eq:Z-exact-abel}
    Z = \int dx_i K^{\rm flat}_{\Sigma_m} (x_i, \dots, ) K^{\rm flat}_{\Sigma_n}(x_i, \dots, ) \dots\,. 
\end{equation}

Let us now go back to the matrix model (\ref{multi-m.m.}) obtained from the localization to constant $F$ argument. By introducing an auxiliary integration variable $\Phi_{\Sigma_m}$ for each
block $\Sigma_m$, we can rewrite (\ref{multi-m.m.}) as 
\begin{equation}
  S = \sum_{\{\Sigma_m\}} \left (\imath (\sum_{i \in \p \Sigma_m} X_{i}, \Phi_{\Sigma_m})  -
    \frac{g^2 A_{\Sigma_m}}{4} \Phi_{\Sigma_m}^2 \right)
\label{first-order-m.m.}
\end{equation}
Here $(,)$ denotes the
positive bilinear form on the Lie algebra $\g$ of $G$ which replaces `$\tr$' in the general case. It is normalized such that the
long roots $\alpha$ have norm $(\alpha, \alpha) = 2$. We have also omitted the orientation factors $s_i^{(m)}=\pm 1$ for simplicity.\footnote{We assume canonical orientation of the boundary $\partial \Sigma_m$ with respect to $\Sigma_m$, and if necessary reabsorb the signs due to orientation into the Wilson loop insertions $\tr_{R_i}e^{\pm X_i}$.} While $X_i \in \g$, it is natural to assume that $\Phi_{\Sigma_m}$ takes values in the dual $\g^{*}$
to the Lie algebra $\g$. To proceed, we can integrate over adjoint $G$-orbits for
each variable $X_i \in \g$ and $\Phi_{\Sigma_m} \in g^{*}$ using Itzykson-Zuber
formula so that the matrix model integral  over $X_i$'s and
$\Phi_{\Sigma_m}$'s reduces to the Cartan $\h$ and its dual $\h^*$. 

Let $X$ be  in the $G$-orbit of  $x_i \in \h$, 
and let $\Phi$ be in the $G$-orbit of $\Lambda \in h^{*}$. 
The Itzykson-Zuber formula for $X \in \g, \Phi \in \g^{*}$ and
integration $Dg $ over the group $G$ with the invariant measure 
reads 
\begin{equation}
\label{eq:Cartan-blocks}
  \int_{ g \in G} Dg e^{  ( g^{-1} X g , \Phi ) } = \frac{ \sum_{w \in W} (-1)^{w}
    e^{(w(x),\Lambda)}}{\Delta(x) \Delta(\Lambda)}
\end{equation}
The matrix model measure $\int [dX_i]$ reduces to $\int dx_i$ with the
measure 
\begin{equation}
  \int [dX_i] = \int dx_i \Delta(x_i)^2
\end{equation}
Since each $x_i$ appears in two blocks touching the same boundary, the
factors $\Delta(x_i)^2$ in the measure cancel the denominators
$\Delta(x_i)$ in the Cartan blocks (\ref{eq:Cartan-blocks}). The factor
$\Delta(\Lambda_{\Sigma_m})^2$ in the measure coming from $\int [d \Phi_{\Sigma_m}]$ combines with the denominators $\Delta(\Lambda_{\Sigma_m})$ in (\ref{eq:Cartan-blocks})
and gives the overall power $\Delta(\Lambda_{\Sigma_m})^{2 - |\p \Sigma_m|}$, that is
exponents 
$1, 0, -1$ for disk, cylinder and triunion respectively. 
Hence, after reduction to the abelian integrations, 
the matrix model partition function with action (\ref{first-order-m.m.}) (with
possibly Wilson loops inserted)  can be written in terms of the blocks 
 \begin{equation}
\label{eq:MM-block}
  K_3^{\rm flat, m.m.}  = \int_{\h^{*}} d \Lambda
\left(  \frac{1}{\Delta(\Lambda)}
\prod_{i=1}^{3}  \left(  \sum_{w_i \in W}
 (-1)^{w_i} e^{ (\Lambda, w_i(x_i))} \right) 
\right) e^{ - \frac{g^2 A_{\rm tri}}{4}  \Lambda^2}
\end{equation}
and similarly for disk and cylinder. 
Notice that under the substitution $\Lambda = \lambda + \rho$, 
the  matrix model blocks (\ref{eq:MM-block}) 
are exactly the same as the exact 2d YM blocks (\ref{eq:building-abel}), except that the summation over the subset of dominant weights $P^{+} \subset \h^{*}$
in the weight lattice $P \subset \h^{*}$ is
replaced by the integration over $\h^{*}$.
Using arguments based on the Poisson resummation or Euler-Maclaurin formula one can see that indeed the matrix model block (\ref{eq:MM-block}), given by the integral, is the
perturbative approximation to the exact 2d YM block
(\ref{eq:building-abel}), given by the sum. To support this general
proof, in appendix we have computed $\langle WWW\rangle$ from the matrix
model and  compared the result to the zero-instanton sector of the exact partition function for gauge group $U(1)$ and $U(2)$.

\section{Gaussian three-matrix model in the planar limit}
\subsection{Catalan numbers}
\begin{figure}
  \centering
  \includegraphics[width=4cm]{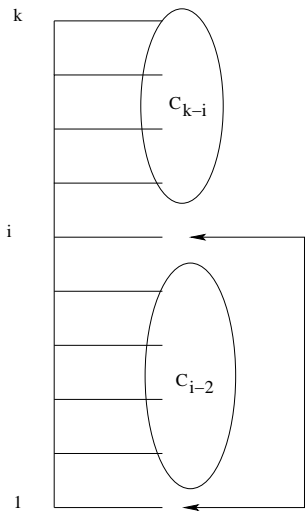}
  \caption{In this recursion we sum over the label $i$ with which the leg ``1'' is contracted.}
\label{fig:1point}
\end{figure}

For start, we recall the usual combinatorial computation for the
 generating function of Catalan numbers, which count the number
 of planar diagrams for the one-point correlation function 
$\frac{1}{N} \tr \la X^k \ra$ in the Gaussian matrix model. By $c(k)$ we
denote the number of such diagrams, so that $c(0) = 1, c(1) =0, c(2) = 1, \dots$.
The numbers  $c(k)$ satisfy the recursion on fig. \ref{fig:1point}
\begin{equation}
  c(k) = \sum_{i=2}^k c(i-2) c(k-i) \label{eq:rec1}.
\end{equation}
Let $f(z)$ be the generating function 
\begin{equation}
  f(z) = \sum_{k=0}^{\infty} c(k) z^{-k-1}.
\end{equation}
From the recursion relation (\ref{eq:rec1}), the generating function
$f(z)$ satisfies
\begin{equation}
  f^2(z) = z f(z) - c_0
\end{equation}
with $c_0 = 1$.
Hence the solution with asymptotics $f(z) = z^{-1} + \dots$ as $z \to \infty$ is 
\begin{equation}
  f(z) = \frac{z - \sqrt{z^2 - 4}}{2}.
\end{equation}
Of course, the generating function 
\begin{equation}
f(z;\lambda) = \lambda^{-1/2} f(\lambda^{-1/2} z)  = 
\frac{ z-
    \sqrt{z^2-4\lambda}}{2 \lambda}
\end{equation}
is actually the planar resolvent
\begin{equation}
  f(z,\lambda) = \la \frac{1}{N} \tr \frac{1}{z - X}\ra_{\mathrm{planar}}
\end{equation}
in the Gaussian matrix model 
\begin{equation}
  Z = \int DX \exp (- \frac{N}{2 \lambda} \tr X^2).
\end{equation}

\subsection{One-point resolvent with external legs}

\begin{figure}
  \centering
  \includegraphics[width=10cm]{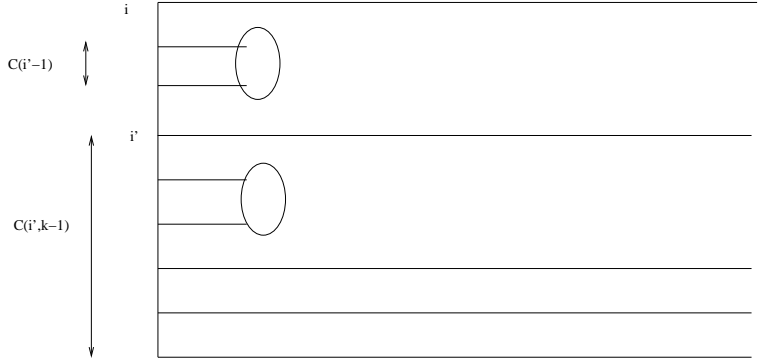}
  \caption{stripe diagrams}
\label{fig:stripe}
\end{figure}

First we compute the number of planar diagrams of the following shape
(fig \ref{fig:stripe}). 
Consider a half-stripe $I \times \BR^{+}$ where an interval $I$ is
vertical and $\BR^{+}$
is a half-line extending to the right, and we are given a set P of $i+1$ points
points placed on $I$ and labelled consecutively from ``0'' to ``i''.

We define a planar diagram with
$k$ external legs as a configuration where $k$ points from the set P are
connected to the positive infinity by horizontal halflines, moreover, 
these $k$ points must include the point ``0'' and ``i'', and the
remaining $i+1-k$ points are connected pairwise by the internal
propagators such that diagram can be drawn on the half-stripe without
intersections.  Let $c_{\mathrm{stripe}}(i,k)$ be the number of such diagrams. 

Then
\begin{equation}
  c_{\mathrm{stripe}}(i,k) = \sum_{i'=0}^{i} c_{\mathrm{stripe}}(i',k-1) c(i-i'-1)
\end{equation}
and
\begin{equation}
  c_{\mathrm{stripe}}(i,1) = \delta_{i0}.
\end{equation}
Let $w(z)_k$ be generating function for $c_{\mathrm{stripe}}(i,k)$ given by
\begin{equation}
  \label{eq:w}
  w(z)_k = \sum_{i=0}^{\infty} c_{\mathrm{stripe}}(i,k) z^{-i}.
\end{equation}
The recursion implies $w(z)_k=w(z)_{k-1}f(z)$, and hence
\begin{equation}
  w(z)_k = f(z)^{k-1}.
\end{equation}
\newcommand{\ccycl}{c_{\mathrm{cycl}}}

Now, let $\ccycl(n,k)$ be the number of planar diagram on a
cylinder $S^1 \times \BR^{+}$ with the set $P$ consisting of $n$
cyclicly labelled points
%$1 \dots n$
$0\dots n-1$
positioned on the circle $S^1$ and $k$ external cyclicly
 labelled legs connecting
$k$ points from $P$ with the infinity with the condition that the legs
are cyclicly oriented. In particular, 
$\ccycl(n,n) = n$ since the are $n$ diagrams obtained by cyclic shifts
from the diagram in which point ``$i$'' in $P$ connects by leg ``$i$'' to
infinity. 

We can deduce $\ccycl(n,k)$ from $c_{\rm stripe}(i,k)$ by summing over the
position of the leg ``$1$'' and ``$k$''. More explicitly, suppose that the $k$ external legs are connected to points on the circle with labels between ``$0$" and ``$i$" ($0\le i\le n-1$), with the external legs ``$1$" and ``$k$" corresponding to points ``$0$" and ``$i$" respectively. Then the number of diagrams is $c_{\rm stripe}(i,k) c(n-i-1)$. Summing over all possible values of ``$i$", and including an overall factor of $n$ by cyclic symmetry due to the arbitrary position of external leg ``$1$", we get 
\begin{equation}
\label{eq:cycl}
  \ccycl(n,k) = n \sum_{i=0}^{n-1} c_{\rm stripe}(i,k) c(n-i-1)\,.
\end{equation}

Consider the generating function for $\ccycl(n,k)$, that is 
the generating
function of the one-point diagrams with $k\geq 1$ external legs 
\begin{equation}
  W(z)_k = \sum_{n=0}^{\infty} \ccycl(n,k) z^{-n-1}
\end{equation}
From (\ref{eq:cycl}) we get
\begin{equation}
 W(z)_k = -\p_{z} (w_k(z) f(z)) = -\p_z f^{k}(z).
\end{equation}
As a check, from this generating function we can extract for instance $\ccycl(1,1)=1,\ccycl(3,1)=3,\ccycl(4,2)=8,\ldots$ which can be seen to correctly count the corresponding number of planar diagrams.

\subsection{Two-point function}
To count two-point planar diagrams we need to glue two effecive vertices
$W_k(z)$ and sum over $k$. The product $W_k (z_1) W_k(z_2)$ overcounts
the number of two point diagrams by factor of $k$ because on a cylinder
we have cyclic symmetry which shifts labelling of the $k$ propagators.
Therefore, in the sector with $k$ propagators connecting the two-points
we get the generating function
\begin{equation}
  W(z_1,z_2)_k = \frac{1}{k} W(z_1)_k W(z_2)_k.
\end{equation}
Now consider the two-matrix Gaussian matrix model with matrices $X_{a}, a =
1,2$
and the propagators 
\begin{equation}
  \la (X_{a})^i_j (X_{b})^k_l \ra = \lambda_{ab} \delta^i_l \delta^k_j
\end{equation}
After we rescale by the propagators $\lambda_{ab}$ and sum over all $k\geq 1$ we get the two-point resolvent
\begin{equation}
  W(z_1,z_2;\lambda_{ab}) = \p_{z_1,z_2} \sum_{k=1}^{\infty}
 \frac{1}{k} (f(z_1;\lambda_{11})f(z_2;\lambda_{22}))^k \lambda_{12}^k
\end{equation}
and finally 
\begin{equation}
  W(z_1,z_2;\lambda_{ab}) = -\p_{z_1,z_2} \log (1 - \lambda_{12}
  f_1 f_2)=\la \tr \frac{1}{z_1-X_1}\, \tr \frac{1}{z_2-X_2} \ra_{\rm conn}
\end{equation}
where $f_a \equiv f(z_a, \lambda_{aa})$. It is easy to check that this agrees with the resolvent derived in \cite{Eynard:1997qu} and used in \cite{Giombi:2009ms},\cite{Giombi:2009ds} to obtain exact predictions for Wilson loops and local operators. In particular, Laplace transforming the above resolvent on $z_2$ and setting $\lambda_{11}=0$ (since we want no propagator from the local operator to itself), one finds the following result
for the correlator of a Wilson loop and a local operator
\begin{equation}
\langle \tr X_1^{J_1} \tr e^{ s_2 X_2}\rangle_{\rm conn}=J_1 \left(\frac{\lam_{12}}{\sqrt{\lam_{22}}}\right)^{J_1} I_{J_1}(2s_2\sqrt{\lam_{22}})\,.
\label{W0-MM}
\end{equation}
which upon inserting the appropriate values of $\lam_{12},\lam_{22}$ gives the result obtained in  \cite{Giombi:2009ds}, see eq. (\ref{WO-2mm-res}) below.

\subsection{Three-point function}
\begin{figure}
   \centering
   \includegraphics[width=8cm]{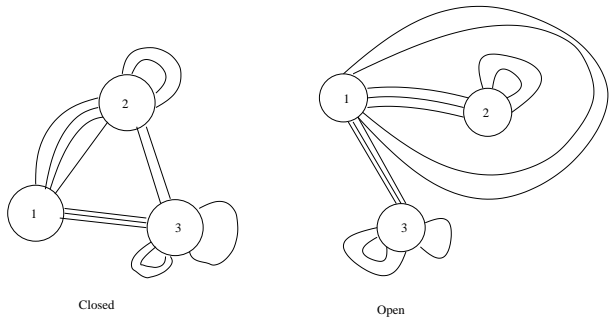}
\caption{In these examples, in the 
closed diagram $k_{12} = 4, k_{23} = 2, k_{13} = 3$, in the
  open diagram $k_{12} = 3$, $k_{13} = 3$, $k_{23} = 0$.}
\label{fig:open-closed}
 \end{figure}

 When computing the three-point planar connected function we shall
 distinguish between two possible global topologies of the diagrams. We
 call them closed and open type (see fig. \ref{fig:open-closed}).

\subsubsection{Closed type diagrams} 

The only difference with the reasoning used to compute the two-point
function, is that on
three-punctured sphere we do not have cyclic symmetry
like on a cylinder. Therefore,
the total number of three-point diagrams is the product of $W(z_i)_{k_a}$:
\begin{equation}
  W(z_1,z_2,z_3) = \prod_{a=1}^{3} W(z_a)_{k_a}
\end{equation}
where $k_{ab}, 1\leq a < b \leq 3, a,b=1\dots 3$ is the number of propagators by
which operator $a$ is connected with operator $b$, and
 $k_1 = k_{12} + k_{23}, k_{2} = k_{23} + k_{12},
k_{3}=k_{13}+k_{23}$.
Now we will sum over all sectors and also we include dependence on the
propagators. Let $\lambda_{ab}$ be the propagator between matrices
$X_a$ and $X_b$. In the sector with $k_{12} = \frac{1}{2} (k_1 + k_2
-k_3)$ propagators we get a factor $\lambda_{12}^{k_{12}}$, etc. 
Then the resolvent for closed type planar diagrams topology is
\begin{equation}
  W(z_1,z_2,z_3; \lambda_{ab})_{k_{12},k_{23},k_{13} \geq 1}=  -\p_{z_1,z_2,z_3} \sum_{k_{12}\geq 1,
    k_{23} \geq 1, k_{13} \geq 1} (\lambda_{12} f_1 f_2)^{k_{12}}
(\lambda_{23} f_2 f_3)^{k_{23}} (\lambda_{13} f_1 f_3)^{k_{13}}
\end{equation}
After the summation we get 
\begin{equation}
\  W(z_1,z_2,z_3; \lambda_{ab})_{k_{12},k_{23},k_{13} \geq 1} = -\p_{z_1,z_2,z_3}
 \frac{ \lambda_{12} \lambda_{23} \lambda_{13} ( f_1  f_2  f_3)^2}{
(1 -  \lam_{12} f_1  f_2)(1 - \lam_{23} f_2 f_3)( 1-  \lam_{13} f_1 f_3)} \label{eq:closed}
\end{equation}
with 
\begin{equation}
  f_a \equiv  f(z_a;\lambda_{aa}).
\end{equation}

\subsubsection{Open topology}
Now we consider the open topology diagrams (see
fig. \ref{fig:open-closed}). Let $k_{12} \geq 1, k_{23} \geq 1, k_{23}
  =0$.

This case is different from the triangle topology because we have extra
diagrams with internal propagators  connecting legs $i$ and $j$ of operator
``1'' such that in between $i$ and $j$ there could be a bunch 
of propagators connecting $O_1$ to $O_2$. 

The total number of diagrams at operator $O_1$ with two bunches
consisting 
of $k_{12}$ and $k_{13}$ external legs is then
\newcommand{\ccycltwo}{c_{\mathrm{cycl2}}}
\begin{equation}
 \ccycltwo(n_1;k_{12},k_{13})=  n_1 \sum_{i=0}^{n_1 - 1} \sum_{j=0}^{n_1 - i -2}(n_1 - 1 - i - j) c_{\mathrm{stripe}}(i,k_{12})
  c_{\mathrm{stripe}}(j,k_{13})  c(n_1 - 2 - i - j)
\end{equation}
In this sum $i =i_{k_{12}}- i_{1}$ and $j = j_{13} - j_{1}$ where
$i_{p}$ denotes the leg in vertex ``1'' connected to the propagator
$p=1\dots k_{12}$ in the first bunch, and $j_{p}$ denotes the leg in
vertex ``1'' connected to the propagator $p=1\dots k_{13}$ in the second
bunch. 
The generating function
\begin{equation}
  W(z_1)_{k_{12},k_{13}} = \sum c_{cycl2}(n_1;k_{12},k_{13}) z_1^{-n_1 -1}
\end{equation}
is 
\begin{equation}
    W(z_1)_{k_{12},k_{13}} = \p_{z_1} ( w(z_1)_{k_{12}} w(z_1)_{k_{13}} (\p_z f(z_1)))
\end{equation}
which simplifies to
\begin{equation}
    W(z_1)_{k_{12},k_{13}} =\frac{1}{k_{12}+k_{13} -1} \p_{z_1}^2 ( f_1^{k_{12}+k_{13}-1}).
\end{equation}

Combining $W(z_1)_{k_{12},k_{13}}$ with the vertices $W(z_2)_{k_{12}}$ and
$W(z_3)_{k_{13}}$ from operators $O_2$ and $O_3$ we get the 
generating function for diagrams with the open 
connected topology in the sector with $k_{23} = 0$ and $k_{12} \geq 1, k_{13} \geq 1$:
\begin{equation}
  W(z_1,z_2,z_3; \lambda)_{k_{23}=0} = \p_{z_1}^2 \p_{z_2} \p_{z_3}
\sum_{k_{12} =1}^{\infty} \sum_{k_{13}=1}^{\infty} \frac{1}{k_{12} +
  k_{13} - 1} f_1^{k_{12}+k_{13}-1}
f_2^{k_{12}} f_3^{k_{13}} \lambda_{12}^{k_{12}} \lambda_{13}^{k_{13}}
\end{equation}
We can sum the series using
\begin{equation}
  \sum_{i=1}^{\infty} \sum_{j=1}^{\infty} \frac{1}{i + j -1} x^i y^j  =
\frac{ xy}{x - y} \log \frac{ 1 - y}{1 - x}
\end{equation}
which can be derived from the series expansion of the integral
\begin{equation}
  \int_0^1 dt \, \frac{ xy}{( 1- tx)(1  - ty)}  = \frac{ xy}{x - y} \log \frac{ 1 - y}{1 - x}.
\end{equation}
So we finally get
\begin{equation}
    W(z_1,z_2,z_3; \lambda_{ab})_{k_{23}=0}  =
\p_{z_1}^2 \p_{z_2} \p_{z_3}  \frac{ \lambda_{12} \lambda_{13} f_2 f_3}{
\lambda_{12} f_2  - \lambda_{13} f_3} \log \frac { 1- \lambda_{13} f_1
f_3}{ 1- \lambda_{12} f_1 f_2}. \label{eq:open}
\end{equation}

\subsubsection{Complete three-point planar resolvent}

The complete three-point planar resolvent in the Gaussian three-matrix model is given by the sum of 
the the resolvent for the closed topology (\ref{eq:closed}) and 
three possible sectors with the open topology (\ref{eq:open}).
The final result for the planar connected three-point resolvent is then
%\begin{multline}
\begin{equation}
\begin{aligned}
    W(z_1,z_2,z_3; \lambda_{ab})&=N \la \tr \frac{1}{z_1-X_1} \tr \frac{1}{z_2-X_2}\tr\frac{1}{z_3-X_3}\ra_{\rm conn} \\ 
&= -\p_{z_1}\p_{z_2}\p_{z_3}
  \frac{ \lambda_{12} \lambda_{23} \lambda_{13} ( f_1  f_2  f_3)^2}{
 (1 -  \lam_{12}f_1  f_2)(1 -  \lam_{23}f_2 f_3)( 1- \lam_{13} f_1 f_3)}  \\
 &+ \p_{z_1}^2 \p_{z_2} \p_{z_3}  \frac{ \lambda_{12} \lambda_{13} f_2 f_3}{
 \lambda_{12} f_2  - \lambda_{13} f_3} \log \frac { 1- \lambda_{13} f_1
 f_3}{ 1- \lambda_{12} f_1 f_2}  \\
&+  \p_{z_1} \p_{z_2}^2 \p_{z_3}  \frac{ \lambda_{12} \lambda_{23} f_1 f_3}{
 \lambda_{12} f_1  - \lambda_{23} f_3} \log \frac { 1- \lambda_{23} f_2
 f_3}{ 1- \lambda_{12} f_2 f_1}  \\
&+ \p_{z_1} \p_{z_2} \p_{z_3}^2 \frac{ \lambda_{23} \lambda_{13} f_1 f_2}{
 \lambda_{23} f_2  - \lambda_{13} f_1} \log \frac { 1- \lambda_{13} f_1
 f_3}{ 1- \lambda_{23} f_2 f_3}
 \label{W3MM}
\end{aligned}
\end{equation}
with $f_a=\frac{z_a-\sqrt{z_a^2-4\lam_{aa}}}{2\lam_{aa}}, a=1,2,3$. As a test, one can verify that if one sets all propagators to be equal, $\lambda_{ab}=\lambda$, then (\ref{W3MM}) reduces as it should to the known 3-point resolvent in the Gaussian one matrix model (see e.g. \cite{Alexandrov:2003pj}). As a further test, we have also computed $\la \tr X_1^{k_1}\tr X_2^{k_2} \tr X_3^{k_3}\ra$ from a direct Feynman diagram calculation at some fixed low values of $k_1,k_2,k_3$, and verified that the result agrees with the series expansion of (\ref{W3MM}).

Note that (\ref{W3MM}) is the resolvent for a general Gaussian 3-matrix model, and hence, once the appropriate $\lambda_{ab}$ are plugged in, it encodes the result for all possible correlators $\langle WWW\rangle$, $\langle WWO\rangle$, $\langle WOO\rangle$ and $\langle OOO\rangle$ in our supersymmetric subsector. The 3-point function of local operators is a particularly simple case (and of course is well known by independent means, since the $O$'s are chiral primaries), so we derive it here as a simple test of our general result for the resolvent. Recall that for three local operators, the 3-matrix model is such that
\begin{equation}
\lambda_{aa}=0\,, \qquad \lambda_{ab} = \frac{\lambda}{16\pi^2}\equiv \lambda_{OO} \quad {\rm for}\,\, a\neq b.
\end{equation}
Then, (\ref{W3MM}) simplifies to
\begin{eqnarray}
W(z_1,z_2,z_3;\lambda_{ab})&=&\partial_{z_1}\partial_{z_2}\partial_{z_3} 
\frac{\lambda_{OO}^2(z_1 z_2+z_2 z_3+z_1 z_3-2\lambda_{OO})}
{(\lambda_{OO}-z_1 z_2)(\lambda_{OO}-z_2 z_3)(\lambda_{OO}-z_1 z_3)}\cr 
&=&\sum_{J_1=1}^{\infty}\sum_{J_2=1}^{\infty}\sum_{k=0}^{{\rm min}(J_1,J_2)} \frac{\lambda_{OO}^{J_1+J_2-k}J_1 J_2 (J_1+J_2-2k)}{z_1^{J_1+1}z_2^{J_2+1}z_3^{J_1+J_2-2k+1}}\,
\end{eqnarray}
from which can read off the 3-point function (we omit the position dependence on the operators, since the correlator does not depend on it)
\begin{equation}
\begin{aligned}
&\langle O_{J_1}O_{J_2}O_{J_3}\rangle =\frac{1}{N} J_1 J_2 J_3 \left(\frac{\sqrt{\lambda}}{4\pi}\right)^{J_1+J_2+J_3}\\
&{\rm for}\,\,|J_1-J_2|\le J_3 \le J_1+J_2 \,,\,\,\,\,\,\,\, J_1+J_2+J_3={\rm even}  
\label{OOO}
\end{aligned}
\end{equation}
and zero otherwise. It is easy to check that this is the expected result for the three-point function of chiral primaries of the type we consider here.\footnote{The result in (\ref{OOO}) is for the operators normalized as $O_J(x)=\tr (x^i \Phi^i+i \Phi^4)^J$. The 3-point function for operators with unit normalized 2-point function is obtained multiplying (\ref{OOO}) by the factor $(i\frac{2\pi}{\sqrt{\lambda}})^{J_1+J_2+J_3}\frac{1}{\sqrt{J_1 J_2 J_3}}$.}

\subsubsection{Exact $<WOO>$ correlator and strong coupling limit}
We now use the exact planar result to derive the exact prediction for the $\langle W O_{J_1}O_{J_2}\rangle$ correlator. It will be first convenient to go to an exponential generating function by taking the Laplace transform of the 3 point resolvent on $z_1,z_2,z_3$. Let us define for convenience 
$$W(z_1,z_2,z_3; \lambda_{ab})\equiv \p_{z_1}\p_{z_2}\p_{z_3}w(f_1,f_2,f_3;\lambda_{ab})$$
where the function $w(f_1,f_2,f_3;\lambda_{ab})$ can be easily read off from (\ref{W3MM}). Then
\begin{equation}
N \langle \tr e^{s_1 X_1} \tr e^{s_2 X_2} \tr e^{s_3 X_3} \rangle_{\rm conn} =
\oint \frac{dz_1 dz_2 dz_3}{(2\pi i)^3} e^{s_1 z_1+s_2 z_2 +s_3 z_3}\p_{z_1}\p_{z_2}\p_{z_3}w(z_1,z_2,z_3;\lambda_{ab})
\end{equation}
where the contours are large circles enclosing the origin. It is convenient to make the change of variables $z_a=\lambda_{aa}f_a+1/f_a$, $\p_{z_a}=\frac{f_a^2}{\lambda_{aa}f_a^2-1}\p_{f_a}$, so that $f(z_a)=f_a$, and the integral turns into a contour integral over $f_1,f_2,f_3$
\begin{equation}
\begin{aligned}
N\langle \tr e^{s_1 X_1} \tr e^{s_2 X_2} \tr e^{s_3 X_3} \rangle_{\rm conn} =
-\oint \frac{df_1 df_2 df_3}{(2\pi i)^3}&e^{s_1 (\lam_{11}f_1+\frac{1}{f_1})+s_2 (\lam_{22}f_2+\frac{1}{f_2}) +s_3 (\lam_{33}f_3+\frac{1}{f_3})}\times\\
&\times \p_{f_1}\p_{f_2}\p_{f_3}w(f_1,f_2,f_3;\lambda_{ab})
\end{aligned}
\end{equation}
where the contours are small circles around the origin. Using the contour integral representation of the modified Bessel function
\begin{equation}
I_k(x)=\oint \frac{dz}{2\pi i}e^{\frac{x}{2}(z+\frac{1}{z})}z^{k-1}=\sum_{n=0}^{\infty}\frac{\left(\frac{1}{2}x\right)^{2n+k}}{n!(n+k)!}
\label{Bessel-contour}
\end{equation}
one finds
\begin{equation}
\begin{aligned}
&N\langle \tr e^{s_1 X_1} \tr e^{s_2 X_2} \tr e^{s_3 X_3} \rangle_{\rm conn} =\sum_{i,j,k=1}^{\infty}\Bigg{[}
\left(\frac{\lam_{12}}{\sqrt{\lam_{11}\lam_{22}}}\right)^{i}
\left(\frac{\lam_{23}}{\sqrt{\lam_{22}\lam_{33}}}\right)^{j}
\left(\frac{\lam_{13}}{\sqrt{\lam_{11}\lam_{33}}}\right)^{k}(i+k)(i+j)(j+k)\times\\
&\qquad\qquad\qquad\qquad\qquad
\times I_{i+k}(2s_1\sqrt{\lam_{11}})I_{i+j}(2s_2\sqrt{\lam_{22}})I_{j+k}(2s_3\sqrt{\lam_{33}})\\
&+\left(\frac{\lam_{12}}{\sqrt{\lam_{11}\lam_{22}}}\right)^{i}
\left(\frac{\lam_{13}}{\sqrt{\lam_{11}\lam_{33}}}\right)^{k}i\, k\, (i+k+2j-2)
I_{i+k+2j-2}(2s_1\sqrt{\lam_{11}})I_{i}(2s_2\sqrt{\lam_{22}})I_{k}(2s_3\sqrt{\lam_{33}})\\
&+\left(\frac{\lam_{12}}{\sqrt{\lam_{11}\lam_{22}}}\right)^{i}
\left(\frac{\lam_{23}}{\sqrt{\lam_{22}\lam_{33}}}\right)^{k}i\, k\, (i+k+2j-2)
I_{i}(2s_1\sqrt{\lam_{11}})I_{i+k+2j-2}(2s_2\sqrt{\lam_{22}})I_{k}(2s_3\sqrt{\lam_{33}})\\
&+\left(\frac{\lam_{23}}{\sqrt{\lam_{22}\lam_{33}}}\right)^{i}
\left(\frac{\lam_{13}}{\sqrt{\lam_{11}\lam_{33}}}\right)^{k}i\, k\, (i+k+2j-2)
I_{k}(2s_1\sqrt{\lam_{11}})I_{i}(2s_2\sqrt{\lam_{22}})I_{i+k+2j-2}(2s_3\sqrt{\lam_{33}})\Bigg{]}\,.
\label{WWW3MM}
\end{aligned}
\end{equation}
So far the result is completely general, since we have not specified the propagators $\lambda_{ab}$. For example, plugging in the appropriate values corresponding to the 3-matrix model (\ref{S-WWW}), the expression (\ref{WWW3MM}) gives the exact prediction for the $\langle WWW\rangle$ correlator, and it is then straightforward to study for example its strong coupling limit. We will not pursue this further here, and focus on $\langle WOO\rangle$ in the following.  

To obtain the exact $W O_{J_1} O_{J_2}$ correlator, we should plug in the $\lambda_{ab}$'s coming from (\ref{eq:OWOaction}) or (\ref{eq:S}), and then extract from (\ref{WWW3MM})  $\langle \tr X_1^{J_1} \tr e^{s_2 X_2} \tr X_3^{J_2} \rangle_{\rm conn}$, by picking up the appropriate power in $s_1,s_3$.\footnote{We leave for now an arbitrary $s_2$ parameter. The singly wound Wilson loop corresponds to $s_2=1$.} 
In fact, the result simplifies considerably if we remember the structure of the particular 3-matrix models we want to solve. They are such that
\begin{equation}
\lam_{11}=\lam_{33}=0
\end{equation}
since we do not allow propagators from a local operator to itself. Then taking the limit $\lam_{11},\lam_{33}\rightarrow 0$ of (\ref{WWW3MM}) and extracting the term proportional to $s_1^{J_1}s_3^{J_2}$ we get
\begin{equation}
\begin{aligned}
&N\langle \tr X_1^{J_1} \tr e^{s_2 X_2} \tr X_3^{J_2} \rangle_{\rm conn} =J_1 J_2 \Bigg{[}
\sum_{k=1}^{\rm{min}(J_1,J_2)-1}\frac{\lam_{12}^{J_1-k}\lam_{23}^{J_2-k}\lam_{13}^k}{\lam_{22}^{\frac{J_1+J_2}{2}-k}} (J_1+J_2-2k)I_{J_1+J_2-2k}(2s_2\sqrt{\lam_{22}})\\
&\qquad \qquad \qquad
+\Theta(J_1-J_2-1)\frac{\lam_{12}^{J_1-J_2}\lam_{13}^{J_2}}{\lam_{22}^{\frac{J_1-J_2}{2}}}(J_1-J_2)I_{J_1-J_2}(2s_2\sqrt{\lam_{22}})\\
&\qquad \qquad \qquad +\sum_{k=1}^{\infty}\frac{\lam_{12}^{J_1}\lam_{23}^{J_2}}{\lam_{22}^{\frac{J_1+J_2}{2}}}(J_1+J_2+2k-2)I_{J_1+J_2+2k-2}(2s_2\sqrt{\lam_{22}})\\
&\qquad \qquad \qquad
+\Theta(J_2-J_1-1)\frac{\lam_{23}^{J_2-J_1}\lam_{13}^{J_1}}{\lam_{22}^{\frac{J_2-J_1}{2}}}(J_2-J_1)I_{J_1-J_2}(2s_2\sqrt{\lam_{22}})\Bigg{]}\,.
\end{aligned}
\label{WOO-final-long}
\end{equation}
where $\Theta(x)$ is the unit step function (with the convention $\Theta(0)=1$). We can simplify this further into the following form
\begin{equation}
\begin{aligned}
&\langle \tr X_1^{J_1} \tr e^{s_2 X_2} \tr X_3^{J_2} \rangle_{\rm conn}=\frac{1}{N}\,J_1 J_2 \left(\frac{\lam_{12}}{\sqrt{\lam_{22}}}\right)^{J_1}\left(\frac{\lam_{23}}{\sqrt{\lam_{22}}}\right)^{J_2}\times\\
&\times \left[\sum_{k=1}^{\rm{min}(J_1,J_2)}\rho^k (J_1+J_2-2k)I_{J_1+J_2-2k}(2s_2\sqrt{\lam_{22}})
+\sum_{k=1}^{\infty}(J_1+J_2+2k-2)I_{J_1+J_2+2k-2}(2s_2\sqrt{\lam_{22}})\right]
\label{WOO-final}
\end{aligned}
\end{equation}
where we have defined
$$
\rho=\frac{\lam_{22}\lam_{13}}{\lam_{12}\lam_{23}}\,.
$$
Eq. (\ref{WOO-final}) is the final result for the correlator $\langle W O O  \rangle_{4d}$ computed from the Gaussian 3-matrix model. To specialize to the two possible topologies of  Fig. \ref{WOO-S2}, one should insert the appropriate values of $\lambda_{ab}$ given in (\ref{open-chain-propagators})-(\ref{closed-chain-propagators}), i.e. 
\begin{equation}
\lambda_{12}=i\lambda \frac{A_2}{A^2}\,,\quad\,\, \lambda_{23}=-i\lambda \frac{A_1}{A^2}\,,\quad\,\,\lambda_{13}=\frac{\lambda}{A^2}\,,\quad\,\,\lambda_{22}=\lambda \frac{A_1A_2}{A^2}
\label{lam-open}
\end{equation}
for the open chain $OWO$ topology  and
\begin{equation}
\lambda_{12}=i\lambda \frac{A_2}{A^2}=\lambda_{23}\,,\quad\,\,\lambda_{13}=\frac{\lambda}{A^2}\,,\quad\,\, \lambda_{22}=\lambda \frac{A_1A_2}{A^2}
\label{lam-closed}
\end{equation}
for the closed chain $OOW$ topology. 

We can now study various strong coupling limits of this result. Let us first look at the limit
\begin{equation}
\lambda \rightarrow \infty\,,\qquad J_1,J_2\ll \sqrt{\lambda}
\label{limit1}
\end{equation}
Assuming that $J_1+J_2$ is even (if it is odd, a similar analysis goes through with minor modifications), we may rewrite the second sum in (\ref{WOO-final}) as
(we set $s_2=1$)
\begin{eqnarray}
%\begin{aligned}
&&\sum_{k=1}^{\infty}(J_1+J_2+2k-2)I_{J_1+J_2+2k-2}(2\sqrt{\lam_{22}})=
\sum_{k=1}^{\infty}(2k)I_{2k}(2\sqrt{\lam_{22}})-\sum_{k=1}^{\frac{J_1+J_2}{2}-1}(2k)I_{2k}(2\sqrt{\lam_{22}})\cr
&&=\sqrt{\lam_{22}} I_1(2\sqrt{\lam_{22}})
-\sum_{k=1}^{\frac{J_1+J_2}{2}-1}(2k)I_{2k}(2\sqrt{\lam_{22}})\,.
\label{sum-rewrite}
%\end{aligned}
\end{eqnarray}
Plugging in the contour integral representation (\ref{Bessel-contour}) of the Bessel function, performing the finite sum and evaluating the large $\lambda$ behavior by saddle point analysis, one finds that
\begin{equation}
\sum_{k=1}^{\frac{J_1+J_2}{2}-1}(2k)I_{2k}(2\sqrt{\lam_{22}})\simeq \frac{1}{4}(J_1+J_2)(J_1+J_2-2)
\frac{e^{2\sqrt{\lam_{22}}}}{\sqrt{4\pi\sqrt{\lam_{22}}}}\qquad \lambda\rightarrow \infty
\label{sum2}
\end{equation}
In particular, we see that the second term in (\ref{sum-rewrite}) is subleading compared to the first one (recall that $I_1(x)\simeq e^x/\sqrt{2\pi x}$ at large $x$). By a similar analysis, one can see that the first sum in (\ref{WOO-final}) is of the same order as (\ref{sum2}), so the leading behavior of the correlator in the limit (\ref{limit1}) is 
\begin{eqnarray}
\langle \tr X_1^{J_1} \tr e^{X_2} \tr X_3^{J_2} \rangle_{\rm conn}\simeq \frac{1}{N}\,J_1 J_2 \left(\frac{\lam_{12}}{\sqrt{\lam_{22}}}\right)^{J_1}\left(\frac{\lam_{23}}{\sqrt{\lam_{22}}}\right)^{J_2}\, \sqrt{\lam_{22}} \frac{e^{2\sqrt{\lam_{22}}}}{\sqrt{4\pi\sqrt{\lam_{22}}}}
\end{eqnarray}
Normalizing by the Wilson loop expectation value
\begin{equation}
\langle \tr e^{ X_2}\rangle = \frac{N}{\sqrt{\lam_{22}}}I_1(2\sqrt{\lam_{22}})
\end{equation}
we find
\begin{eqnarray}
\frac{\langle \tr X_1^{J_1} \tr e^{ X_2} \tr X_3^{J_2} \rangle_{\rm conn}}{\langle \tr e^{ X_2}\rangle}
&&\simeq\frac{1}{N^2}\left[\sqrt{\lam_{22}}\,J_1  \left(\frac{\lam_{12}}{\sqrt{\lam_{22}}}\right)^{J_1}\right]
\left[\sqrt{\lam_{22}}\,J_2\left(\frac{\lam_{23}}{\sqrt{\lam_{22}}}\right)^{J_2}\right]\cr
&&\simeq \frac{\langle \tr X_1^{J_1} \tr e^{ X_2}\rangle_{\rm conn}}{\langle \tr e^{ X_2}\rangle}\,
\frac{\langle \tr e^{ X_2} \tr X_3^{J_2} \rangle_{\rm conn}}{\langle \tr e^{X_2}\rangle}\,,
\label{factorized}
\end{eqnarray}
where in the last step we have used the two-point function (\ref{W0-MM}). We will see in the next section that this factorized structure is indeed reproduced in string theory. 

We can now move on to the more interesting strong coupling limit
\begin{equation}
\lambda\rightarrow \infty\,,\qquad J_1/\sqrt{\lambda}={\rm fixed}\equiv {\cal J}_1\,,\qquad J_2\ll \sqrt{\lambda}
\label{limit2}
\end{equation}
For comparison to string theory, it is convenient to look at the normalized correlator
\begin{equation}
\frac{\langle \tr X_1^{J_1} \tr e^{X_2} \tr X_3^{J_2} \rangle_{\rm conn}}{\langle \tr X_1^{J_1} \tr e^{ X_2}\rangle_{\rm conn}}
\end{equation}
where the two point function at the denominator is given in (\ref{W0-MM}).
To extract the strong coupling limit, we can employ as usual the contour integral representation of the modified Bessel function and apply the saddle point analysis. The term in square bracket in (\ref{WOO-final}) gives, after performing the summation and keeping the leading terms in the $J_1\gg J_2$ limit
\begin{equation}
\oint \frac{dz}{2\pi i}\left(J_1\rho\,\frac{\rho^{J_2} z^{-J_2}-z^{J_2}}{\rho-z^2}+J_1\frac{z^{J_2}}{1-z^2}\right)z^{J_1-1}e^{\sqrt{\lam_{22}}(z+\frac{1}{z})}
\label{z-repr}
\end{equation} 
Writing (see eq. (\ref{lam-open}),(\ref{lam-closed}))
$$
J_1=\sqrt{\lambda}{\cal J}_1\,,\qquad \lam_{22}=a^2\lam/4\,,\quad  a^2\equiv 4A_1A_2/A^2
$$  
the saddle point is the solution of
\begin{equation}
{\cal J}_1\frac{1}{z}+\frac{a}{2}(1-\frac{1}{z^2})=0\,.
\label{saddle-eq1}
\end{equation}
The solution which gives the dominant exponential at large $\lambda$ is 
\begin{equation}
z_{*}=\sqrt{1+\frac{{\cal J}_1^2}{a^2}}-\frac{{\cal J}_1}{a}\,.
\label{saddle-eq2}
\end{equation}
Evaluating (\ref{z-repr}) on this saddle point and dividing by the $\langle O_{J_1}W\rangle$ correlator we then obtain (note that the exponential and the ``1-loop'' correction around the saddle point cancel out in the normalization by the two point function)
\begin{equation}
\frac{\langle \tr X_1^{J_1} \tr e^{X_2} \tr X_3^{J_2} \rangle_{\rm conn}}{\langle \tr X_1^{J_1} \tr e^{ X_2}\rangle_{\rm conn}}
\simeq \frac{J_2}{N}\left(\frac{\lam_{23}}{\sqrt{\lam_{22}}}\right)^{J_2}\sqrt{\lambda}{\cal J}_1\left(\rho\,\frac{\rho^{J_2} z_{*}^{-J_2}-z_{*}^{J_2}}{\rho-z_{*}^2}+\frac{z_{*}^{J_2}}{1-z_{*}^2}\right)\,.
\end{equation}
To compare to string theory, it will be convenient to insert also the normalization factor (\ref{norm-O}) for the local operators, i.e. 
${\cal N}_J=\left(-i\frac{2\pi}{\sqrt{\lam}}\right)^{J}\frac{1}{\sqrt{J}}$.
Using the explicit $\lambda_{ab}$ in (\ref{lam-open})-(\ref{lam-closed}), we get for the open chain (for which one has $\rho=1$)
\begin{equation}
\frac{\langle \tr X_1^{J_1} \tr e^{X_2} {\cal N}_{J_2}\tr X_3^{J_2} \rangle_{\rm conn}}{\langle \tr X_1^{J_1} \tr e^{ X_2}\rangle_{\rm conn}}\simeq \frac{1}{N}\,(-1)^{J_2}2^{-J_2}\sqrt{J_2\lam}\left(\frac{A_1}{A_2}\right)^{J_2/2}\,{\cal J}_1 \frac{z_{*}^{-J_2}}{1-z_{*}^2}\,.
\label{WOO-open}
\end{equation}
and for the closed chain (for which $\rho=-\frac{A_1}{A_2}$)
\begin{equation}
\frac{\langle \tr X_1^{J_1} \tr e^{X_2} {\cal N}_{J_2}\tr X_3^{J_2} \rangle_{\rm conn}}{\langle \tr X_1^{J_1} \tr e^{ X_2}\rangle_{\rm conn}}\simeq \frac{1}{N}\,2^{-J_2}\sqrt{J_2\lam}\left(\frac{A_2}{A_1}\right)^{J_2/2}\,{\cal J}_1
\left(\frac{A_1}{A_2}\,\frac{\left(-\frac{A_1}{A_2}\right)^{J_2} z_{*}^{-J_2}-z_{*}^{J_2}}{\frac{A_1}{A_2}+z_{*}^2}+\frac{z_{*}^{J_2}}{1-z_{*}^2}\right)\,.
\label{WOO-closed}
\end{equation}
These are the predictions of the 3-matrix model in the limit (\ref{limit2}). In the next section we will compare them to the result one obtains from semiclassical strings in $AdS_5\times S^5$.

Finally, let us look at the limit in which all operators are in the semiclassical regime
\begin{equation}
\lambda\rightarrow \infty, \qquad J_1/\sqrt{\lambda}={\rm fixed}\equiv {\cal J}_1,\qquad 
J_2/\sqrt{\lambda}={\rm fixed}\equiv {\cal J}_2\,.
\label{limit3}
\end{equation}
Let us assume without loss of generality that $J_1\ge J_2$. Passing to the contour integral representation of the Bessel functions and performing the sums in (\ref{WOO-final}), one gets the following structure
\begin{equation}
N\langle \tr X_1^{J_1} \tr e^{X_2} \tr X_3^{J_2} \rangle_{\rm conn}=
\oint \frac{dz}{2\pi i} \left[f_1(z)\rho^{\sqrt{\lam}{\cal J}_2}z^{\sqrt{\lam}({\cal J}_1-{\cal J}_2)}
+f_2(z)z^{\sqrt{\lam}({\cal J}_1+{\cal J}_2)}\right]e^{\frac{a\sqrt{\lam}}{2}(z+\frac{1}{z})}
\end{equation}
where $f_1(z), f_2(z)$ have at most a linear dependence in $\sqrt{\lambda}$. The saddle point analysis at large $\lambda$ then gives the behavior
\begin{eqnarray}
N\langle \tr X_1^{J_1} \tr e^{X_2} \tr X_3^{J_2} \rangle_{\rm conn} 
&&\simeq c_1 e^{\sqrt{\lambda}\left(\sqrt{a^2+({\cal J}_1-{\cal J}_2)^2}
+({\cal J}_1-{\cal J}_2)\log\left(\sqrt{1+\frac{({\cal J}_1-{\cal J}_2)^2}{a^2}}-\frac{{\cal J}_1-{\cal J}_2}{a}\right)+{\cal J}_2\log\rho\right)}\cr
&&+c_2 e^{\sqrt{\lambda}\left(\sqrt{a^2+({\cal J}_1+{\cal J}_2)^2}
+({\cal J}_1+{\cal J}_2)\log\left(\sqrt{1+\frac{({\cal J}_1+{\cal J}_2)^2}{a^2}}-\frac{{\cal J}_1+{\cal J}_2}{a}\right)\right)}\,,
\label{WOO-bigJJ}
\end{eqnarray}
We see that there are two saddle points, both with positive exponents, contributing at large $\lambda$. It can be seen that the saddle point in the first line is the dominant one (recall that we are assuming ${\cal J}_1\ge {\cal J}_2$). These saddle points should correspond to semiclassical string solutions describing a Wilson loop worldsheet with two ``fattened'' spikes. It would be interesting to find those solutions. 

\section{String theory computations}
\subsection{Semiclassical $<WO>$ correlator}

For finite R-charge $J$, it is well known that the correlator $\langle WO_J \rangle$ at strong coupling can be computed (as originally done in \cite{Berenstein:1998ij}) via the exchange of a supergravity mode between the string worldsheet dual to the Wilson loop and the operator insertion point at the AdS boundary. Here we are interested instead in the limit of large R-charge $J \sim \sqrt{\lambda} \gg 1$. In this limit the correlator is effectively described by a new string solution where the propagator for the supergravity mode has ``fattened'' into an infinite spike \cite{Zarembo:2002ph}.

In \cite{Zarembo:2002ph}, the string solution corresponding to $\langle W_{\mbox{\tiny 1/2-BPS}}O_J \rangle$ in the large R-charge limit was derived ($W_{\mbox{\tiny 1/2-BPS}}$ is the well-known 1/2-BPS circular Wilson loop). Here we obtain a generalization of that solution to the lower supersymmetric case $\langle W_{\mbox{\tiny 1/4-BPS}}O_J \rangle$. The Wilson loop $W_{\mbox{\tiny 1/4-BPS}}$ is the quarter-BPS operator of \cite{Drukker:2006ga} (which is also a special case of the infinite family of loops on $S^2$ of \cite{Drukker:2007qr}). It corresponds to a circular contour in space-time and couples to 3 scalar fields which we take to be $\Phi_1,\Phi_2,\Phi_3$. The chiral primary we consider is given by $\tr(\Phi_3+i \Phi_4)^J$.

In the original definition of \cite{Drukker:2006ga}, the 1/4-BPS Wilson loop is specified by the circular contour (which we can take to have say unit radius) $x^{\mu}(s)=(\cos s ,\sin s ,0,0)$, and couples to the scalar field combination
\begin{equation}
-\cos\theta_0 \cos s\, \Phi_1-\cos\theta_0 \sin s\, \Phi_2+\sin \theta_0\, \Phi_3\,.
\end{equation}
The choice $\theta_0=\pi/2$ corresponds to the 1/2-BPS Wilson loop. By a conformal transformation, one can map the space-time contour $x^{\mu}(s)$ to be a latitude on $S^2$ at angle $\theta_0$: this gives an equivalent definition of $W_{\mbox{\tiny 1/4-BPS}}$ which fits inside the infinite family of loops of \cite{Drukker:2007qr}. To derive the string solution below, it will be more convenient to consider the original definition of \cite{Drukker:2006ga} and do the conformal transformation to the $S^2$ setup in a second step.

The string solution of interest lies on a $AdS_3\times S^3$ subspace of $AdS_5\times S^5$, on which we take the metric
\begin{equation}
ds^2=\frac{1}{Z^2}\left(dZ^2+dR^2+R^2d\phi^2\right)+d\vartheta^2+\sin^2\vartheta d\varphi^2+\cos^2\vartheta d\psi^2\,.
\end{equation}
Defining embedding coordinates on $S^3$ as 
\begin{equation}
\Theta^A=(-\sin\vartheta \cos\varphi,-\sin\vartheta \sin\varphi,\cos\vartheta\cos\psi,\cos\vartheta\sin\psi), 
\end{equation}
the local operator we wish to insert corresponds to the spherical harmonic $\left(\Theta^3+i\Theta^4\right)^J=(\cos\vartheta)^J e^{i J \psi}$.

We will assume worldsheet conformal gauge, with worldsheet coordinates $(\tau,\sigma)$, and take the following rotationally symmetric ansatz
\begin{equation}
\begin{aligned}
&Z=Z(\tau), \quad R=R(\tau), \quad \vartheta=\vartheta(\tau)\, \quad \psi=\psi(\tau)\\
&\phi=\varphi=\sigma\, \qquad 0< \sigma\le 2\pi\,\quad 0< \tau< \infty\,.
\end{aligned}
\end{equation}
The string action specialized to this ansatz reads
\begin{equation}
S=\frac{\sqrt{\lambda}}{2}\int d\tau \left[ \frac{1}{Z^2}\left(\dot Z^2+\dot R^2+R^2\right)+\dot\vartheta^2 +\sin^2\vartheta+\cos^2\vartheta \dot\psi^2 \right]\,
\label{string-action}
\end{equation}
where the dot indicates $\tau$-derivative.

The solution dual to the 1/4-BPS Wilson loop, without local operator insertion, is given by
\begin{equation}
\begin{aligned}
&Z=\tanh\tau\,\qquad R=\frac{1}{\cosh\tau}\\
&\sin\vartheta=\frac{1}{\cosh(\tau\pm \tau_0)}\,,\quad \psi=0\,,\qquad \tanh\tau_0=\sin\theta_0\,.
\label{quarter-string}
\end{aligned}
\end{equation}
In these coordinates, the Wilson loop at the boundary sits at $\tau=0$. The choice of $\pm$ sign correspond to the existence of two inequivalent solutions, one stable and one unstable, which are dual to the same Wilson loop \cite{Drukker:2006ga} (see also \cite{Giombi:2009ms}). For simplicity in the following we will focus on the stable solution, corresponding to the choice of $+$-sign, but everything below has a counterpart for the corresponding unstable solution.

Let us now derive the solution corresponding to the insertion of the local operator $\tr(\Phi_3+i \Phi_4)^J$, with $J/\sqrt{\lambda}={\rm fixed}\equiv {\cal J}$. The insertion of the local operator deforms the above solution by creating an infinite spike which terminates at $\tau\rightarrow \infty$. The angle $\psi$ also assumes a non-trivial profile. The boundary conditions for the various fields are as follows \cite{Zarembo:2002ph}. At $\tau=0$, they are controlled by the Wilson loop, i.e.
\begin{equation}
Z(0)= 0\,,\quad R(0) = 1\,,\quad \vartheta(0)= \frac{\pi}{2}-\theta_0\,,\quad \psi(0)=0\,.
\end{equation}
At $\tau\rightarrow \infty$, the boundary condition are dictated by the presence of the local operator and are given by \cite{Zarembo:2002ph}
\begin{equation}
\log Z(\tau)\rightarrow {\cal J}\tau \,,\quad R(\tau) \rightarrow 0\,,\quad \vartheta(\tau)\rightarrow 0 \,,\quad \psi(\tau)\rightarrow i {\cal J} \tau \,, \qquad {\rm at}\,\,\, \tau\rightarrow \infty\,. 
\label{bc-infinity}
\end{equation}
In particular the boundary condition on $\psi$ corresponds to the fact that we have rotation on a great circle of $S^5$ with the appropriate angular momentum $J$ (after analytic continuation to Minkowski signature on the worldsheet).

The equations of motion following from the above action
\begin{equation}
\begin{aligned}
&\partial_{\tau} \left(\frac{\dot Z}{Z}\right)+\frac{1}{Z^2}\left(\dot R^2+R^2\right)=0\,\qquad
\partial_{\tau} \left(\frac{\dot R}{Z^2}\right)-\frac{R}{Z^2}=0\,,\\
&\ddot \vartheta-\sin\vartheta\cos\vartheta (1-\dot\psi^2)=0\,,\qquad \partial_{\tau} ( \cos^2\vartheta\, \dot\psi)=0\,,
\end{aligned}
\end{equation}
can be solved because they admit the following integrals of motion
\begin{equation}
\begin{aligned}
&\frac{\dot Z^2}{Z^2}+\frac{1}{Z^2}\left(\dot R^2-R^2\right)=\epsilon\\
&\frac{R \dot R }{Z^2}+\frac{\dot Z}{Z}=\kappa\\
&\cos^2\vartheta\, \dot\psi=p_{\psi}\\
&\dot\vartheta^2-\sin^2\vartheta+\frac{p_{\psi}^2}{\cos^2\vartheta}=-\epsilon\,.
\label{conserved}
\end{aligned}
\end{equation}
where $\epsilon$, $\kappa$, $p_{\psi}$ are constants. The first and last integrals of motion are just the ``energies'' corresponding to the $AdS_3$ and $S^3$ parts of the lagrangian, while the second one follows from invariance under dilatation. The fact that the constants in the first and last line have equal and opposite values is a consequence of the Virasoro constraints. One can see that the boundary conditions in (\ref{bc-infinity}) further fix $p_{\psi}=i {\cal J}$, $\epsilon={\cal J}^2$, $\kappa={\cal J}$.

\begin{figure}
\begin{center}
\includegraphics[width=60mm]{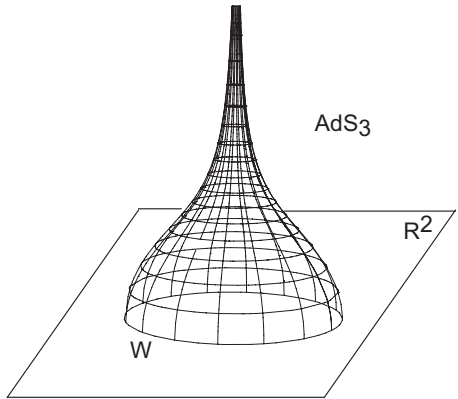}
\parbox{13cm}{\caption{$AdS_3$ part of the string solution (\ref{sol}) describing the correlator $\langle W O_J\rangle$ for large R-charge $J\sim \sqrt{\lambda}$. The infinite spike corresponds to the local operator insertion.}
\label{WO-R2}}
\end{center}
\end{figure}

To integrate the first two lines of (\ref{conserved}), it is convenient to introduce the variables (see also \cite{Drukker:2005cu})
\begin{equation}
u=\frac{R}{Z}\,,\qquad v=\log \sqrt{R^2+Z^2}
\end{equation}
in terms of which we have
\begin{equation}
\begin{aligned}
\dot u^2-u^4-(1+\epsilon)u^2=\epsilon-\kappa^2\,,\qquad \quad
\dot v = \frac{\kappa}{1+u^2}\,.
\end{aligned}
\end{equation}
These equations, as well as the last two lines of (\ref{conserved}), can be integrated in a straightforward way in terms of elliptic integrals. In fact, the solution becomes elementary once one imposes that $\epsilon=\kappa^2=-p_{\psi}^2={\cal J}^2$.
Going back to the $R,Z$ variables, the solution with the correct boundary conditions turns out to be
\begin{eqnarray}
% &&Z=e^{{\cal J} \tau} \left(\sqrt{1+{\cal J}^2}
%   \tanh\left({\sqrt{1+{\cal J}^2} \tau+\tau_0}\right)-{\cal J}\right)\,,\cr
&&Z = e^{{\cal J} \tau} \frac { \sinh ( \sqrt{ 1 + {\cal J}^2} \tau)}
{ \cosh ({ \sqrt{ 1 + { \cal J}^2} (\tau + \tau_0)})} \\
&&R=\frac{\sqrt{1+{\cal J}^2} e^{{\cal J}
    \tau}}{\cosh\left({\sqrt{1+{\cal J}^2}(\tau+\tau_0)}\right)}\,,
\quad \sinh(\sqrt{1+{\cal J}^2}\tau_0) = {\cal J}\cr
&& \label{sol}
 \sin\vartheta = \frac{ \sqrt{1 + {\cal J}^2}}{ \cosh ( \sqrt{ 1  + {\cal
       J}^2}(\tau + \tau_1))}, \quad \sinh(\sqrt{1+{\cal J}^2}\tau_1)=\frac{ \sqrt{
    {\cal J}^2+\sin^2\theta_0}}{\cos\theta_0} \\
% &&
% \sin\vartheta=\frac{1+{\cal J}^2}{\cosh\left({\sqrt{1+{\cal
%           J}^2}(\tau+\tau_0)}\right)+\frac{{\cal J}^2}{2}
%   e^{-\sqrt{1+{\cal J}^2}(\tau+\tau_0)}}\,,\\
 &&e^{i\psi}=e^{-{\cal J}\tau}\,
 \frac{\sqrt{1+\frac{{\cal J}^2}{\sin^2\theta_0}}-\frac{{\cal J}}{\sin\theta_0}}{\sqrt{1+\frac{{\cal J}^2}{\cos^2\vartheta}}-\frac{{\cal J}}{\cos\vartheta}}
 \,. \nonumber
 \end{eqnarray}
One can see that for $\theta_0=\pi/2$ (i.e. $\tau_1=\infty$) one recovers the solution of Zarembo \cite{Zarembo:2002ph}, while for ${\cal J}=0$ we recover the solution dual to the 1/4-BPS Wilson loop (\ref{quarter-string}). A depiction of the $AdS_3$ part of the solution is given in Figure \ref{WO-R2}. 

\begin{figure}
\begin{center}
\includegraphics[width=60mm]{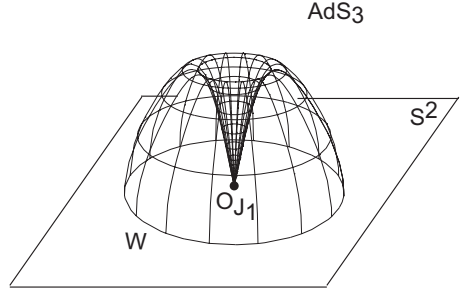}
\parbox{13cm}{\caption{The string solution describing the semiclassical $\langle W O_J\rangle$ correlator after the conformal transformation (\ref{conf-trans}). The Wilson loop is a latitude on $S^2=\partial AdS_3$, here depicted as a plane, and the spike ends on the north pole of $S^2$, where the local operator is inserted.}
\label{WO-S2}}
\end{center}
\end{figure}

It is not difficult to write down the conformal transformation that maps the solution found above to the setup in which the Wilson and local operators are inserted on a $S^2$ subspace of the boundary \cite{Drukker:2007qr}\cite{Giombi:2009ds}. Consider $AdS_4$ with metric
\begin{equation}
ds^2=\frac{1}{z^2}\left(dz^2+dr^2+r^2d\phi^2+dx_3^2\right)\,.
\end{equation}
The transformed string solution lies on the $AdS_3$ slice defined by $z^2+r^2+x_3^2=1$, and is given by
\begin{equation}
\begin{aligned}
&z=\frac{2 c_0 Z}{1+c_0^2(Z^2+R^2)}\,\\
&r=\frac{2 c_0 R}{1+c_0^2(Z^2+R^2)}\,\\
&x_3=\frac{-1+c_0^2(Z^2+R^2)}{1+c_0^2(Z^2+R^2)}\,,\qquad c_0=\cot\frac{\theta_0}{2}
\label{conf-trans}
\end{aligned}
\end{equation}
where $R$, $Z$ correspond to the solution in (\ref{sol}). The angle $\phi=\sigma$ as well as the $S^5$ part of the solution are unchanged. One can see that at $\tau=0$ the solution ends on a latitude on $S^2$ of angle $\theta_0$. Notice that now $\tau=\infty$ is also a point on the $AdS$ boundary, since $z(\tau=\infty)=0$. Indeed we see that after the conformal transformation the spike at $\tau=\infty$ ends on the north pole of the $S^2$ (i.e. $z=0$, $x_3=1$, $r=0$), which is then interpreted as the insertion point of the local operator in the boundary theory, see Figure \ref{WO-S2}. Recall that in the $S^2$ setup the interesting local operators have the form $\tr\left(x^i\Phi^i(x)+i \Phi^4\right)^J$, so an operator inserted at the north pole indeed corresponds to $\tr\left(\Phi^3(x)+i \Phi^4\right)^J$, as used in the derivation of the string solution above.

To obtain the strong coupling prediction for the correlator $\langle W_{\theta_0}O_J \rangle$, we need to evaluate the string action on the solution, supplemented by a boundary contribution at infinity which follows from the insertion of the local operator at $\tau=\infty$. After using in (\ref{string-action}) the values of the integrals of motion, the ``bulk'' action is
\begin{equation}
S_{\mbox \tiny bulk}=\sqrt{\lambda}\int d\tau \left(\frac{r^2}{z^2}+\sin^2\vartheta\right)=\sqrt{\lambda}\int d\tau\left(\frac{R^2}{Z^2}+\sin^2\vartheta\right)\,.
\label{S-bulk}
\end{equation}
This has a well-understood divergence at $\tau=0$, which in this case we can simply eliminate by considering the normalized correlator $\langle W_{\theta_0}O_J \rangle/\langle W_{\theta_0} \rangle$, i.e. by subtracting the value of the action at ${\cal J}=0$. After performing the $\tau$-integral 
\begin{equation}
\begin{aligned}
&\int_0^{\infty}d\tau \left(\frac{R^2}{Z^2}-\frac{R^2}{Z^2}\Big{|}_{{\cal J}=0}\right)=1-\sqrt{1+{\cal J}^2}\\
&\int_0^{\infty}d\tau \left(\sin^2\vartheta-\sin^2\vartheta|_{{\cal J}=0}\right)
=\sin\theta_0-1+\sqrt{1+{\cal J}^2}-\sqrt{{\cal J}^2+\sin^2\theta_0}
\end{aligned}
\end{equation}
we obtain
\begin{equation}
S_{\mbox \tiny bulk}-S_{\mbox \tiny bulk}|_{{\cal J}=0}=\sqrt{\lambda}\left(\sin\theta_0-\sqrt{{\cal J}^2+\sin^2\theta_0}\right)\,.
\end{equation}
The boundary contribution at infinity 
originates from the insertion of the vertex operator for the local operator at the north pole, which, omitting subleading factors in the large $J=\sqrt{\lambda}{\cal J}$ limit, is essentially given by 
\begin{equation}
\left(\frac{z}{z^2+r^2+(x_3-1)^2}\right)^J (\cos\vartheta)^J e^{i J \psi} =
\left(\frac{c_0}{2}Z\right)^J (\cos\vartheta)^J e^{i J \psi}.
\label{Vop}
\end{equation}
This corresponds to adding to the action the boundary term
\begin{equation}
S_{\mbox \tiny bdy}=-\sqrt{\lambda} \left({\cal J} \log Z(\infty)+i {\cal J} \psi(\infty)+{\cal J} \log \cos\vartheta(\infty)\right)-\sqrt{\lambda}{\cal J}\log\frac{c_0}{2}\,.
\end{equation}
The last term in the first parenthesis vanishes, while the divergences in the first two terms cancel each other leaving the result
\begin{equation}
S_{\mbox \tiny bdy}=-\sqrt{\lambda}{\cal J} \log\left( \sqrt {1+\frac{{\cal J}^2}{\sin^2\theta_0}}-\frac{{\cal J}}{\sin\theta_0}\right)-\sqrt{\lambda}{\cal J}\log\frac{c_0}{2}\,.
\end{equation}
Putting everything together, we then find
\begin{equation}
\frac{\langle W_{\theta_0}{\cal O}_J \rangle}{\langle W_{\theta_0} \rangle}=
\frac{e^{-S_{\mbox \tiny bulk}-S_{\mbox \tiny bdy}}}{e^{-S_{\mbox \tiny bulk}|_{{\cal J}=0}}}=e^{\sqrt{\lambda}\left(\sqrt{{\cal J}^2+\sin^2\theta_0}-\sin\theta_0+{\cal J} \log\left( \sqrt {1+\frac{{\cal J}^2}{\sin^2\theta_0}}-\frac{{\cal J}}{\sin\theta_0}\right)\right)+\sqrt{\lambda}\log\left(\frac{1}{2}\cot\frac{\theta_0}{2}\right)}
\label{string-result}
\end{equation}
We can now compare this to the field theory prediction. The relation to 2d YM implies that the exact Wilson-local correlator should be given by the Gaussian two-matrix model result in eq. (\ref{WO-2mm-res}), see \cite{Giombi:2009ds}. Inserting the values of the couplings $\lam_{12},\lam_{22}$ which can be read off from the action in (\ref{WO-2mm}), as well as the normalization factor (\ref{norm-O}) for the local operator, (\ref{W0-MM}) yields the prediction, for an arbitrary Wilson loop on $S^2$ and local operator inserted at the north pole (or any other point on the upper hemisphere)
\begin{equation}
\frac{\la {\cal O}_J(x_N) W(C)\ra}{\la W(C)\ra}= \frac{1}{2} \sqrt{J \lambda '}  \left(\frac{A_2}{4 A_1}\right)^{\frac{J}{2}} \frac{I_J(\sqrt{\lambda '})}{I_1(\sqrt{\lambda '})}\,
\label{WO-2mm-res}
\end{equation}
where $\lambda '=4A_1A_2/A^2\lambda$. 
Recall that for a latitude $4A_1A_2/A^2=\sin^2\theta_0$ and $A_2/A_1=\cot^2\frac{\theta_0}{2}$. Thus we see that the prefactor $\left(\frac{A_2}{4A_1}\right)^{J/2}=e^{\sqrt{\lambda}{\cal J}\log \frac{1}{2}\sqrt{A_2/A_1}}=e^{\sqrt{\lambda}{\cal J}\log \left(\frac{1}{2}\cot\frac{\theta_0}{2}\right)}$ reproduces the last term in the exponent in (\ref{string-result}). The remaining terms arise from the strong coupling limit of the Bessel function
\begin{equation}
I_J(\sqrt{\lambda}\sin\theta_0)=\oint \frac{dz}{2\pi i} z^{J-1} e^{\frac{1}{2}\sqrt{\lambda}\sin\theta_0(z+\frac{1}{z})}\,.
\end{equation}
The saddle point analysis at large $\lambda$ and large R-charge $J={\cal J}\sqrt{\lambda}$ is the same as in (\ref{saddle-eq1})-(\ref{saddle-eq2}), and the dominant exponent at strong coupling\footnote{There is also a subleading saddle point which corresponds to an unstable string solution, see refs} gives
\begin{equation}
I_J(\sqrt{\lambda}\sin\theta_0)\sim e^{\sqrt{\lambda}\left(\sqrt{{\cal J}^2+\sin^2\theta_0}
+{\cal J}\log\left(\sqrt{1+{\cal J}^2/\sin^2\theta_0}-{\cal J}/\sin\theta_0\right)\right)}\,.
\end{equation}
After dividing by $I_1(\sqrt{\lambda}\sin\theta_0)\sim e^{\sqrt{\lambda}\sin\theta_0}$, we therefore precisely reproduce the string theory prediction (\ref{string-result}).

\subsection{$<WOO>$ correlator}
In this section, we present the string theory computation of the correlator $\langle W_{\theta_0}\,O_{J_1}\, O_{J_2} \rangle$, and we then compare to strong coupling limit of the 3-matrix model prediction. 

\begin{figure}
\begin{center}
\includegraphics[width=100mm]{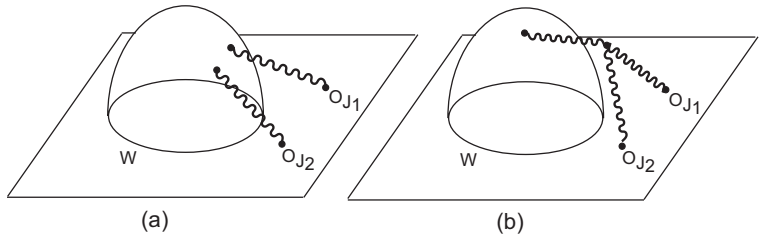}
\parbox{13cm}{\caption{The two diagrams contributing at leading order at large $N$ to $\langle W O_{J_1}O_{J_2}\rangle$ for $J_1,J_2\ll\sqrt{\lambda}$. Fig. (b) is subleading in the large $\lambda$ limit.}
\label{WOO-sm}}
\end{center}
\end{figure}

Let us start with the case in which both local operators are ``light'', i.e. $J_1,J_2 \ll \sqrt{\lambda}$. In this regime, the correlator at strong coupling should be computed by considering the exchange of two supergravity modes between the boundary and the worldsheet describing the 1/4-BPS Wilson loop \ref{quarter-string}. There are two types of contributions, one involving a 3-point interaction in the bulk, and one with two independent propagators ending on distinct points on the worldsheet, see Figure \ref{WOO-sm}. As explained in \cite{Buchbinder:2010ek} in the case of the analogous calculation of 4-point functions of local operators, the diagram involving the 3-vertex is subleading at large $\lambda$, since it involves only one vertex operator inserted on the worldsheet. On the other hand, the diagram with two independent bulk-to-boundary propagators simply factorizes, so we immediately conclude that
\begin{equation}
\frac{\langle W_{\theta_0}\, {\cal O}_{J_1}(x_1)\,{\cal O}_{J_2}(x_2)\rangle }{\langle W_{\theta_0}\rangle}
\simeq \frac{\langle W_{\theta_0}\, {\cal O}_{J_1}(x_1)\rangle}{\langle W_{\theta_0}\rangle}\,\frac{\langle W_{\theta_0}\, {\cal O}_{J_2}(x_2)\rangle}{\langle W_{\theta_0}\rangle}, \qquad \sqrt{\lambda}\gg 1,\, J_1,J_2\ll \sqrt{\lambda}
\end{equation}
Each of the factors on the right-hand-side is the same as computed in \cite{Giombi:2009ds}, in particular it agrees with the two-point function in the two-matrix model. The factorized structure precisely agrees with the 3-matrix model in the relevant limit, see eq. (\ref{factorized}).\footnote{It is easy to see that a similar leading factorization into a product of $\langle W O\rangle$ applies more genereally to the correlator of a Wilson loop and several local operators, as long as the operators are ``light".} It would be interesting to compute the subleading corrections from the diagram in Fig. 8(b), and compare to the corrections to (\ref{factorized}) on the matrix model side.\footnote{For a recent evaluation of the analogous subleading corrections to the 4-point function of two ``heavy" and two ``light" operators, see \cite{Arnaudov:2012xx}.}

\begin{figure}
\begin{center}
\includegraphics[width=80mm]{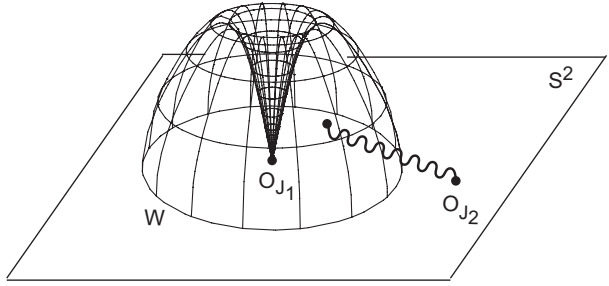}
\parbox{13cm}{\caption{Diagram describing the string theory computation of the correlator $\langle W O_{J_1} O_{J_2}\rangle$ for $J_1\sim \sqrt{\lambda}$ and $J_2\ll J_1$.}
\label{WOO}}
\end{center}
\end{figure}

Next we look at the more interesting case in which one operator is ``heavy'' $J_1\sim \sqrt{\lambda} \gg 1$, while the other is ``light'', i.e. $J_2$ is kept of order one. In this case, we can simply use the string solution for $\langle W_{\theta_0}\,O_{J_1}\rangle$ derived earlier, and compute the amplitude for the exchange of a supergravity mode corresponding to $O_{J_2}$, see Figure \ref{WOO}. The recipe to do this is well understood, see e.g. \cite{Zarembo:2010rr}\cite{Roiban:2010fe} for analogous calculations involving three-point functions of local operators. Here we follow the notations of \cite{Giombi:2009ds}, where the appropriate vertex operator for local operators inserted on $S^2$ was given.
We have
\begin{equation}
\begin{aligned}
&\frac{\langle W_{\theta_0}\,{\cal O}_{J_1}(\vec{x}_N)\, {\cal O}_{J_2}(\vec{x}_0) \rangle}{\langle W_{\theta_0}\,{\cal O}_{J_1}(\vec{x}_N)\rangle}=\int d^2\sigma \left(V_{J_2}^{AdS_5}+V_{J_2}^{S^5}\right)\\
&V_{J_2}^{AdS_5}=\frac{\sqrt{\lambda}}{4\pi N}\frac{J_2+1}{4\sqrt{J_2}}
\left(2 \frac{J_2(J_2-1)}{J_2+1}\frac{1}{z^2}\partial_{\alpha}x^{\mu}\partial_{\alpha}x^{\mu}G_{J_2}(x^{\mu};x_0)
-\frac{4}{J_2+1}\partial_{\alpha}\partial_{\alpha}G_{J_2}(x^{\mu};x_0)\right)Y^{J_2}(\Theta,x_0)\\
&V_{J_2}^{S^5}=-\frac{\sqrt{\lambda}}{4\pi N}\frac{J_2+1}{4\sqrt{J_2}}\, 2J_2\,\partial_{\alpha}\Theta^A\partial_{\alpha}\Theta^A \,G_{J_2}(x^{\mu};x_0) Y^{J_2}(\Theta,x_0)\\
&x^{\mu}=(z,\vec{x})\,,\quad \vec{x}=(r\cos\phi,r\sin\phi,x_3)\,,\quad \\
&\Theta^A=(-\sin\vartheta \cos\varphi,-\sin\vartheta \sin\varphi,\cos\vartheta\cos\psi,\cos\vartheta\sin\psi)\\
&G_{J_2}=\left(\frac{z}{z^2+(\vec{x}-\vec{x}_0)^2}\right)^{J_2}=
\left(\frac{z}{2-2\vec{x}\cdot \vec{x_0}}\right)^{J_2}\,,\qquad
Y^{J_2}=\left(\vec{x}_0\cdot \vec{\Theta}+i \Theta^4\right)^{J_2}\,,
\end{aligned}
\end{equation}
where $x^{\mu},\Theta^A$ correspond to the solution (\ref{sol}), with $z,r,x_3$ related to $R,Z$ in (\ref{sol}) by the conformal transformation (\ref{conf-trans}) (and of course ${\cal J}={\cal J}_1\equiv J_1/\sqrt{\lambda}$). The notation $x_N$ indicates the north pole of $S^2$, where the heavy operator is inserted, while we keep the insertion point of the light operator $x_0\in S^2$ arbitrary. We can choose the parameterization
$$
x_0=(\sin\eta \cos\gamma,\sin\eta \sin\gamma,\cos\eta)\,,\qquad 0\le \eta \le \pi,\quad 0\le \gamma<2\pi\,.
$$
By rotational invariance of the problem, we can always set $\gamma=0$, which we will assume in the following.
We can simplify these expressions further if we integrate by parts the second term in $V_{J_2}^{AdS_5}$. This produces a term proportional to $\partial^2 Y^{J_2}$. Using the classical equations of motion $\partial^2\Theta^A+(\partial_{\alpha}\Theta^B\partial_{\alpha}\Theta^B)\Theta^A=0$, we can rewrite this as 
\begin{equation}
\partial_{\alpha}\partial_{\alpha}Y^{J_2}=\left[J_2(J_2-1)\left(\partial_{\alpha}\log\left(\vec{x}_0\cdot \vec{\Theta}+i \Theta^4\right)\right)^2-J_2 \partial_{\alpha}\Theta^A\partial_{\alpha}\Theta^A\right] Y^{J_2}.
\end{equation}
Then, further using the fact that on the explicit solution $\frac{1}{z^2}\partial_{\alpha}x^{\mu}\partial_{\alpha}x^{\mu}=2\frac{r^2}{z^2}+{\cal J}_1^2$ and $\partial_{\alpha}\Theta^A\partial_{\alpha}\Theta^A=2\sin^2\vartheta-{\cal J}_1^2$, we arrive at
\begin{eqnarray}
%\begin{aligned}
&&\frac{\langle W_{\theta_0}\,{\cal O}_{J_1}(\vec{x}_N)\, {\cal O}_{J_2}(\vec{x}_0) \rangle}{\langle W_{\theta_0}\,{\cal O}_{J_1}(\vec{x}_N)\rangle}=\cr
&&=\frac{\sqrt{J_2 \lambda}}{4\pi N}(J_2-1)
\int d\tau d\sigma \Bigg{[}\frac{r^2}{z^2}-\sin^2\vartheta+{\cal J}_1^2-\left(\partial_{\alpha}\log\left(\vec{x}_0\cdot \vec{\Theta}+i \Theta^4\right)\right)^2 \Bigg{]}G_{J_2} Y^{J_2}\,\cr
&&~~~~~~~~~~~~~~~~~~~
-\frac{\sqrt{J_2 \lambda}}{4\pi N J_2}\int_0^{2\pi}d\sigma  \left[(\partial_{\tau}G_{J_2})Y^{J_2}-G_{J_2}(\partial_{\tau}Y^{J_2})\right]{\Big{|}}_{\tau=0}^{\tau=\infty}
%\end{aligned}
\label{WOO-string}
\end{eqnarray}
The boundary term in the second line is easily evaluated. The contribution from $\tau=0$, where the Wilson loop is inserted, vanishes.\footnote{Recall that $J_2\ge 2$.} On the other hand at $\tau\rightarrow \infty$ we have
\begin{equation}
\begin{aligned}
&G_{J_2}=\left(\frac{e^{-{\cal J}_1\tau}}{c_0(1-\cos\eta)(\sqrt{1+{\cal J}_1^2}-{\cal J}_1)}\right)^{J_2}+\ldots\\
&Y^{J_2}=\left(-\frac{1}{2}(1-\cos\eta)e^{{\cal J}_1\tau}\frac{\sqrt{1+{\cal J}_1^2}-{\cal J}_1}{\sqrt{1+\frac{{\cal J}_1^2}{\sin^2\theta_0}}-\frac{{\cal J}_1}{\sin\theta_0}}\right)^{J_2}+\ldots
\end{aligned}
\end{equation} 
so that 
\begin{equation}
-\frac{\sqrt{J_2 \lambda}}{4\pi N J_2}\int_0^{2\pi}d\sigma  \left[(\partial_{\tau}G_{J_2})Y^{J_2}-G_{J_2}(\partial_{\tau}Y^{J_2})\right]{\Big{|}}_{\tau=0}^{\tau=\infty}=
2^{-J_2} \sqrt{J_2\lambda}{\cal J}_1  
\left(\frac{-\tan\frac{\theta_0}{2}}{\sqrt{1+\frac{{\cal J}_1^2}{\sin^2\theta_0}}-\frac{{\cal J}_1}{\sin\theta_0}}\right)^{J_2}
\label{bdy-term}
\end{equation}
Note in particular that the dependence on the insertion point $x_0$ of the light operator (which enters through the parameter $\eta$, see above) drops out.

The integral in the first line of (\ref{WOO-string}) is more challenging to compute, given the complicated form of the solution.  A limit in which it can be evaluated analytically is $\theta_0=\frac{\pi}{2}$ and ${\cal J}_1\rightarrow 0$ (in this case the calculation essentially reduces to the $\langle W O\rangle$ correlator considered in \cite{Giombi:2009ds}). In particular, in this limit one can explicitly show the independence on the position of the local operator $O_{J_2}$, see appendix B. 
In any case, a numerical evaluation is straightforward for all ranges of the parameters. Keeping in mind the structure of the matrix model result  (\ref{WOO-open})-(\ref{WOO-closed}) which we would like to match, we can simply guess the analytic result and compare it to numerics. We find that the numerical integration agrees with very good precision, over a wide range of the parameters $J_2,{\cal J}_1,\theta_0,\eta$, with the following analytic expression
\begin{eqnarray}
&&\frac{\sqrt{J_2 \lambda}}{4\pi N}(J_2-1)
\int d\tau d\sigma \Bigg{[}\frac{r^2}{z^2}-\sin^2\vartheta+{\cal J}_1^2-\left(\partial_{\alpha}\log\left(\vec{x}_0\cdot \vec{\Theta}+i \Theta^4\right)\right)^2 \Bigg{]}G_{J_2} Y^{J_2}=\cr 
&&=\begin{cases}\frac{2^{-J_2}}{N}\sqrt{J_2\lam}\left(-\tan\frac{\theta_0}{2}\right)^{J_2}\,{\cal J}_1 \frac{z_{*}^{2-J_2}}{1-z_{*}^2}
\qquad \qquad \qquad \qquad \qquad  \qquad \qquad \quad \,\,
\theta_0<\eta\le \pi\\
\frac{2^{-J_2}}{N}\sqrt{J_2\lam}\left(\cot\frac{\theta_0}{2}\right)^{J_2}\,{\cal J}_1
\left(-\frac{\left(-\tan^2\frac{\theta_0}{2}\right)^{J_2} z_{*}^{2-J_2}+\tan^2\frac{\theta_0}{2}z_{*}^{J_2}}{\tan^2\frac{\theta_0}{2}+z_{*}^2}+\frac{z_{*}^{J_2}}{1-z_{*}^2}
\right)\qquad 0<\eta<\theta_0
\end{cases}
\label{bulk-int}
\end{eqnarray}
Here we have used, as in the matrix model calculation, the shorthand notation
$$
z_{*}=\sqrt{1+\frac{{\cal J}_1^2}{\sin^2\theta_0}}-\frac{{\cal J}_1}{\sin\theta_0}
$$
Note that, as a non-trivial result of the integration, we find that the correlator is independent of the insertion point $x_0$, except for a jump when the operator crosses the loop. This is again as expected from the localization to the 2d theory. \footnote{It might be possible to prove the position independence of the correlator for a generic solution in the supersymmetric subsector on $S^2$. One should be able to represent the variation of the integrand under the shift of the position $x_0\in S^2$ of the vertex operator as an exact two-form using the calibration conditions (see the next subsection) and the fact that the
operator $\delta O$, where $\delta$ is a shift along $S^2$, is Q-exact.}

Putting together the integral (\ref{bulk-int}) and the contribution of the boundary term (\ref{bdy-term}), our final result for the 3-point correlator from string theory is 
\begin{eqnarray}
&&\frac{\langle W_{\theta_0}\,{\cal O}_{J_1}(\vec{x}_N)\, {\cal O}_{J_2}(\vec{x}_0) \rangle}{\langle W_{\theta_0}\,{\cal O}_{J_1}(\vec{x}_N)\rangle}=\cr
&&=\begin{cases}
\frac{2^{-J_2}}{N}\sqrt{J_2\lam}\left(-\tan\frac{\theta_0}{2}\right)^{J_2}\,{\cal J}_1 \frac{z_{*}^{-J_2}}{1-z_{*}^2}
\qquad \qquad \qquad \qquad \qquad  \qquad \qquad \quad  \,\,
x_0\in S^{-}\\ 
\frac{2^{-J_2}}{N}\sqrt{J_2\lam}\left(\cot\frac{\theta_0}{2}\right)^{J_2}\,{\cal J}_1
\left(\tan^2\frac{\theta_0}{2}\,\frac{\left(-\tan^2\frac{\theta_0}{2}\right)^{J_2} z_{*}^{-J_2}-z_{*}^{J_2}}{\tan^2\frac{\theta_0}{2}+z_{*}^2}+\frac{z_{*}^{J_2}}{1-z_{*}^2}\right)\qquad x_0\in S^+ \setminus \{x_N\}
\end{cases}
\end{eqnarray} 
where $S^{\pm}$ denote the two regions of $S^2$ singled out by the loop, with the north pole $x_N\in S^+$. Recall that for a loop along a latitude with angle $\theta_0$ we have
\begin{equation}
A_1=4\pi\sin^2\frac{\theta_0}{2},\qquad A_2=4\pi \cos^2\frac{\theta_0}{2},
\end{equation}
and so we see that the result of the string calculation precisely matches the 3-matrix model prediction (\ref{WOO-open})-(\ref{WOO-closed}).

\subsection{Supersymmetry equations}
In this subsection we show that the string solution in (\ref{sol}) dual to the $\langle W O \rangle$ correlator preserves the same supercharges found in the field theory analysis. Moreover we derive a set of first order differential equations which should be satisfied by the string solutions dual to correlators of any number of Wilson loops and local operators. 

Let us start with the $\kappa$-symmetry equation for a fundamental string. Since the relevant solutions lie on $AdS_3\times S^3$, we can restrict the analysis to such subspace of $AdS_5\times S^5$. It will be actually convenient to view $AdS_3\times S^3$ as embedded in $AdS_4\times S^3$ with metric
\begin{equation}
\begin{aligned}
&ds^2=\frac{1}{z^2}dx^i dx^i+z^2(dy^idy^i+(dy^4)^2)\qquad i=1,2,3 \\
&z^{-2}\equiv y^iy^i+(y^4)^2\,.
\label{7d-metric}
\end{aligned}
\end{equation}
The equation defining the relevant $AdS_3\times S^3$ subspace is 
\begin{equation}
x^ix^i+z^2=1\,.
\end{equation}
In what follows, we will denote the coordinates on $AdS_4\times S^3$ as $X^M=(x^i,y^i,y^4)$, $M=1,\ldots,7$. 

In our conventions and in Euclidean signature, the $\kappa$-symmetry equations read
\begin{equation}
i \partial_{\tau}X^M\partial_{\sigma}X^N \Gamma_{MN}\epsilon_{AdS}=\sqrt{h}\epsilon_{AdS}\,.
\end{equation}
Here $\Gamma_{MN}=1/2[\Gamma_M,\Gamma_N]$ and $\Gamma_M$ are curved 7d Dirac matrices, $\epsilon_{AdS}$ is the $AdS_5\times S^5$ Killing spinor and $h$ is the determinant of the worldsheet metric. We assume worldsheet conformal gauge, so that
\begin{equation}
\sqrt{h}=\partial_{\tau}X^M\partial_{\tau}X^N G_{MN}=\partial_{\sigma}X^M\partial_{\sigma}X^N G_{MN}\,,
\end{equation}
where $G_{MN}$ is the metric (\ref{7d-metric}). 

Restricted to $AdS_4\times S^3$ with metric (\ref{7d-metric}), the explicit form of the Killing spinor is
\begin{equation}
\epsilon_{AdS}=\frac{1}{\sqrt{z}}\left(\epsilon_0+z (x^i\Gamma_i-y^i\Gamma_{i+3}-y^4\Gamma_7)\epsilon_1\right)
\equiv \frac{1}{\sqrt{z}}\left(\epsilon_0+\tilde{X}^M\Gamma_M\epsilon_1\right),
\label{killing}
\end{equation}
where for later convenience we have introduced the notation
\begin{equation}
\tilde{X}^M=z(x^i,-y^i,-y^4)\,.
\end{equation}
Note that on the $AdS_3\times S^3$ subspace we have $\tilde{X}^M\tilde{X}_M=1$.
Note also that $\tilde{X}^M$ is orthogonal to all vectors in the tangent space of $AdS_3\times S^3$
\begin{equation}
\tilde{X}^Mp_M=0\qquad \forall p_M\in T(AdS_3\times S^3)
\end{equation}
as follows from differentiating the constraint $x^ix^i+z^2=1$.

In (\ref{killing}) $\epsilon_0$ and $\epsilon_1$ are constant 16 components (Majorana-Weyl) spinors which are directly identified with the Poincare and superconformal supersymmetry parameters in the ${\cal N}=4$ SYM at the boundary. In particular, to impose that the string solution preserves the same supercharges as the dual operators, $\epsilon_0$ and $\epsilon_1$ should satisfy 
\begin{equation}
\begin{aligned}
&i\gamma_{jk}\epsilon_0=\epsilon_{ijk}\rho_i\epsilon_1\,,\qquad 
i\gamma_{jk}\epsilon_1=\epsilon_{ijk}\rho_i\epsilon_0 \\
&i\gamma_i\rho_4\epsilon_0=\rho_i\epsilon_1\,,\qquad 
i\gamma_i\rho_4\epsilon_1=\rho_i\epsilon_0
\end{aligned}
\label{susy1}
\end{equation}
which imply that
\begin{equation}
\begin{aligned}
&(\gamma_{ij}+\rho_{ij})\epsilon_0=0\,,\qquad \gamma_{123}\epsilon_0=\rho_4\epsilon_0\\
&\epsilon_1=-i\rho_{123}\epsilon_0\,.
\end{aligned}
\label{susy2}
\end{equation}
We can collect all these conditions in the following equations, written in terms of curved 7d Dirac matrices
\begin{equation}
i\Gamma_{MN}\epsilon_0=J_{MNP}\Gamma^P \epsilon_1\,,\qquad 
i\Gamma_{MN}\epsilon_1=J_{MNP}\Gamma^P \epsilon_0\,.
\label{Gam-rule}
\end{equation}
Here $J_{MNP}$ is a totally antisymmetric tensor whose components can be read off from (\ref{susy1})-(\ref{susy2}). They are the components of the following 3-form
\begin{eqnarray}
J_{(3)}&=&\frac{1}{3!}J_{MNP}dX^M\wedge dX^N \wedge dX^P\nonumber\\
&=&\frac{1}{2z}\epsilon_{ijk} dx^i \wedge dx^j\wedge dy^k-\frac{z^3}{6}\epsilon_{ijk} dy^i\wedge dy^j\wedge dy^k
-z dx^i\wedge dy^i\wedge dy^4\,.
\end{eqnarray}
Similarly to \cite{Drukker:2007qr}, from $J_{(3)}$ we can construct an almost-complex structure on $AdS_3\times S^3$. Indeed, one can see that
\begin{equation}
J^M_{\ N}\equiv J^M_{\ N P}\tilde{X}^P
\end{equation}
satisfies on $AdS_3\times S^3$
\begin{equation}
J^M_{\ P}J^P_{\ N}=-\delta^M_N+\tilde{X}^M\tilde{X}_N
\end{equation}
and hence it squares to $-1$ when acting on a vector in the tangent space $T(AdS_3\times S^3)$. It is also easy to see that $J^M_{\ N}$ maps tangent vectors to tangent vectors, and so it defines an almost-complex structure on $AdS_3\times S^3$.\footnote{In fact, its components are closely related to those of the almost-complex structure on $AdS_4\times S^2$ constructed in \cite{Drukker:2007qr} for general supersymmetric Wilson loops on $S^3$.} From $J^M_{\ N}$ one can construct as usual a 2-form $J_{(2)}=1/2 J_{MN}dX^M\wedge dX^N$, which however is not closed.

Going back to the $\kappa$-symmetry equations, we start from
\begin{equation}
\left(i\partial_{\tau}X^M\partial_{\sigma}X^N\Gamma_{MN}-\sqrt{h}\right)
(\epsilon_0+\tilde{X}^P\Gamma_P\epsilon_1)=0
\end{equation}
and use the supersymmetry conditions (\ref{Gam-rule}). This gives
\begin{eqnarray}
\left(\partial_{\tau}X^M\partial_{\sigma}X^N\tilde{X}^Q\Gamma_QJ_{MNP}\Gamma^P
-\sqrt{h}\right)\epsilon_0+
\left(\partial_{\tau}X^M\partial_{\sigma}X^N J_{MNP}\Gamma^P-\sqrt{h}\tilde{X}_P\Gamma^P\right)\epsilon_1=0\nonumber\,,
\end{eqnarray}
where we have used that $[\Gamma_{MN},\Gamma_Q]=2G_{NQ}\Gamma_M-2G_{MQ}\Gamma_N$ and that $\tilde{X}_M\partial_{\tau}X^M=\tilde{X}_M\partial_{\sigma}X^M=0$. Writing $\Gamma_Q\Gamma^P=\delta_Q^P+\Gamma_{Q}^{\ P}$ and using again (\ref{Gam-rule}) as well as the last line of (\ref{susy2}) we end up with
\begin{eqnarray}
&&\Bigg{[}\left(\partial_{\tau}X^M\partial_{\sigma}X^N J_{MN}-\sqrt{h}\right)\mathbf{1} \\
&&-\frac{i}{z^3}\left(\partial_{\tau}X^M\partial_{\sigma}X^N J_{MNP}-\sqrt{h}\tilde{X}_P
-i\partial_{\tau}X^M\partial_{\sigma}X^N J_{MNQ}J^{Q}_{\ P}\right)\Gamma^P \Gamma_{456}\Bigg{]}\epsilon_0=0\,.\nonumber
\end{eqnarray}
Since we have already imposed all supersymmetry constraints, the terms multiplying the identity and $\Gamma^P\Gamma_{456}$ should vanish independently. The condition coming from the identity 
\begin{equation}
\begin{aligned}
&\partial_{\tau}X^M\partial_{\sigma}X^N J_{MN}-\sqrt{h}=0
\end{aligned}
\label{id-eq}
\end{equation}
is the statement that the string solution is ``calibrated'' by $J_{(2)}$ (in a generalized sense, since $J_{(2)}$ is not closed), i.e. the worldsheet area is given by 
\begin{equation}
A(\Sigma)=\int_{\Sigma} J_{(2)}\,.
\label{calib}
\end{equation}
Note that the calibration condition is actually not independent from the remaining seven conditions
\begin{equation}
\partial_{\tau}X^M\partial_{\sigma}X^N J_{MNP}-\sqrt{h}\tilde{X}_P
-i\partial_{\tau}X^M\partial_{\sigma}X^N J_{MNQ}J^{Q}_{\ P}=0
\label{susy-eq}
\end{equation}
since (\ref{id-eq}) follows from contracting the above equations by $\tilde{X}^P$ (recall that $\tilde{X}^M\tilde{X}_M=1$ and $J^{Q}_{\ P}\tilde{X}^P=0$). 

The fact that the string solutions are calibrated by $J_{(2)}$ is analogous to the result obtained in \cite{Drukker:2007qr} for the Wilson loops on $S^3$. There is however an important difference in the present case involving mixed correlators of Wilson loops and local operators. If the string solution $X^M(\tau,\sigma)$ were {\it real}, then from 
\begin{equation}
\left(J^M_{\ N}\partial_{\sigma}X^N-\partial_{\tau}X^M\right)^2=2\sqrt{h}
-2J_{MN}\partial_{\tau}X^M\partial_{\sigma}X^N=0
\label{JdX-sq}
\end{equation}
one could conclude that the solution is pseudo-holomorphic with respect to $J$, i.e.
\begin{equation}
J^M_{\ N}\partial_{\sigma}X^N=\partial_{\tau}X^M\,.
\label{J-eq}
\end{equation}
One can also check that if (\ref{J-eq}) holds, then (\ref{susy-eq}) are automatically satisfied.\footnote{To see this, one can note that $J_{(3)}$ satisfies the identity $J_{MNQ}J^{Q}_{\ P}=G_{MP}\tilde{X}_N-G_{NP}\tilde{X}_M
+\frac{1}{3!}\epsilon_{MNPQ_1Q_2Q_3Q_4}\tilde{X}^{Q_1}J^{Q_2Q_3Q_4}$.} So for real solutions, the supersymmetry equations are equivalent to the pseudo-holomorphicity equations. This is the case for correlators of Wilson loops alone, as in \cite{Drukker:2007qr}. However, in the cases involving some local operators, the boundary conditions on the Euclidean worldsheet are such that some angle along $S^5$ is purely imaginary, see (\ref{bc-infinity}) \footnote{While the analysis in the previous section was for $\langle W O \rangle$, a similar boundary condition would apply also in the general case, at each spike corresponding to a local operator.}, so the solution $X^M(\tau,\sigma)$ is not in general real. Therefore, while the solutions are still calibrated (\ref{calib}), the pseudo-holomorphic equations do not follow from (\ref{JdX-sq}), and one should solve the more general supersymmetry equations (\ref{susy-eq}). 

To support the general analysis of this subsection, we have explicitly checked by direct computation that the solution (\ref{sol}) found earlier indeed  satisfies (\ref{susy-eq}), proving that it preserves the correct supersymmetries. On the other hand, it is easy to see that it does not satisfy the pseudo-holomorphicity equations (\ref{J-eq}), unless ${\cal J}=0$. 

It would be interesting to study further the supersymmetry equations (\ref{susy-eq}) and see if they can be used to find new solutions dual to more general $\langle WW\cdots OO\cdots\rangle$ correlators. In particular they could be useful to find the solution dual to $\langle W O_{J_1} O_{J_2}\rangle$ for $J_1,J_2\sim \sqrt{\lambda}$, for which we have an explicit prediction (\ref{WOO-bigJJ}) from the 3-matrix model. We leave this for future work.

\subsection*{Acknowledgments}
We are grateful to J. Gomis, N. Gromov, I. Klebanov and A. Sever for
useful discussions, and especially to R. Roiban and A. Tseytlin for comments on a preliminary draft of this paper. S.G. would like to thank the Aspen Center for Physics and the NSF Grant 1066293 for hospitality during completion of this work. S.G. is supported by Perimeter Institute for Theoretical Physics. Research at Perimeter Institute is supported by the Government of Canada through Industry Canada and by the Province of Ontario through the Ministry of Research $\&$ Innovation.  V.P. gratefully acknowledges support from Institute for Advanced Study, NFS grant PHY-0969448, MSERF 14.740.11.0081, NSh 3349.2012.2 and RFBR grant 10-02-00499.

\appendix 
\section{$<WWW>$ for $U(1)$ and $U(2)$: matrix model vs exact 2d YM partition function}
\subsection{U(1)}
Let $G = U(1)$. Let the holonomies be $U_i = e^{i w_i x_i}$.  Let the Wilson
loops be  $ e^{i w_i x_i}$ where $w_i$ are fixed integer weights. 

Gluing $K_{\rm tri}$ and three $K_{\rm disk}$ and integrating over $x_i$ we get 
\begin{equation}
  \la W_1 W_2 W_3 \ra_{U(1)} =\frac{1}{Z} \sum_{n \in \BZ} e^{ - \frac{1}{4} ( n^2
    a + (n + w_1)^2 a_1 + (n+w_2)^2 a_2 + (n+w_3)^2 a_3)}
\end{equation}
where here $n$ labels the summation index over irreps in the block
$K_{\rm tri}$, and for shortness we absorbed $g^2$ into areas $a,a_i$, where
$a$ is the area of the triunion and $a_i$ are areas of the disks. Now
we will make use of the Poisson resummation identity
\begin{equation}
  \sum_{n \in \BZ} \delta (y - n) = \sum_{m \in \BZ} e^{2 \pi i m y},
\end{equation}
so that 
\begin{equation}
\begin{aligned}
  \la W_1 W_2 W_3 \ra_{U(1)} &=\frac{1}{Z} \int dy \sum_{m \in \BZ} e^{2 \pi i m y} 
 e^{ - \frac{1}{4} ( y^2
    a + (y + w_1)^2 a_1 + (y+w_2)^2 a_2 + (y+w_3)^2 a_3)} \\
&= 
\frac{1}{Z} \sum_{m \in \BZ} e^{ \frac{ (\frac 1 2 a_i w_i + 2 \pi i m)^2}{a+\sum
    a_i} - \frac 1 4 \sum_i w_i^2 a_i}
\label{U1-exact}
\end{aligned}
\end{equation}
where now the meaning of $m$ is labelling 2d magnetic fluxes, or 2d
unstable instantons. The perturbative, or zero-instanton, contribution comes from the $m = 0$ term:
\begin{equation}
\label{eq:U1pert}
  \la W_1 W_2 W_3 \ra_{pert,U(1)}
 = e^{ \frac{ (\sum  a_i w_i  )^2}{4(a+\sum      a_i)} -\frac 1 4 \sum_i w_i^2 a_i}
\end{equation}
If we repeat the localization argument of \cite{Giombi:2009ms} to the
constant curvature configurations as described in section \ref{constant-F}, 
we get the Gaussian 3-matrix model 
\begin{equation}
\label{eq:MMaction}
\la W_1 W_2 W_3 \ra_{pert} =   \int dX_1 dX_2 dX_3 e^{-\frac{  (X_1 + X_2 + X_3)^2}{a}- \sum
    \frac{  X_i^2}{a_i}} W(X_1) W(X_2) W(X_3)
\end{equation}
where $X_i$ take value in the Lie algebra of $G$, and Wilson loops are
defined on the group elements $e^{i X_i} \in G$. 
Not surprisingly, the Gaussian integration of the $U(1)$ matrix model with $W(x_i) = e^{i w_i x_i}$, using the standard formula $\int e^{- \frac 1 2 x
  H x + i J x} = e^{ -\frac 1 2 J H^{-1} J} $, reproduces the
perturbative part of the correlation function from the exact Migdal
formula for $U(1)$ (\ref{U1-exact}). 
\subsection{SU(2)}
We will label irreps of spin $j$ of $SU(2)$ by their dimension $n=2j+1$, so
that $n \in \BZ_{>0}$. 
The characters are 
\begin{equation}
  \chi_n(x) = \frac{ \sin n x}{\sin x}
\end{equation}
The Casimir is 
\begin{equation}
  C_2 = 2j (j+1) = \frac{1}{2} (n^2-1)
\end{equation}
(our Casimir is twice of the standard quantum mechanics textbook
conventions  $\sum_i L_i^2 = j(j+1)$ because of the
normalization  $\tr_F L_1 L_1 = \tr_F L_2 L_2 =\tr_F L_3 L_3 = \frac 1 2$ ). 
The building blocks are
\begin{equation}
  \begin{aligned}
    K_{\rm tri}(x_1,x_2,x_3) = \sum_{n=1}^{\infty} e^{-\frac{ n^2 - 1}{8} a
    }\frac{1}{n} \frac {\sin n x_1}{\sin x_1}  \frac {\sin n x_2}{\sin
      x_2}  \frac {\sin n x_3}{\sin x_3}  \\
K_{\rm disk}(x_i) =\sum_{n_i=1}^{\infty} e^{ - \frac{n_i^2-1}{8} a_i }
n_i \frac{ \sin n_i x_i}{ \sin x_i}
  \end{aligned}
\end{equation}
The Haar measure to integrate over $G = \SU(2)$ is 
\begin{equation}
  \int dx \, \sin^2 x 
\end{equation}
Let the Wilson loops be in the fundamental $W = \frac{1}{2} \tr_{F} U(x)$, i.e.
\begin{equation}
  W_i(x_i) = \cos x_i\,.
\end{equation}
Integration over $x_i$ is trivial
\begin{equation}
  \int \sin n x \, \sin n x_1 cos x_1 = \frac{1}{2} (\delta_{n, n_1 + 1}
  + \delta_{n, n_1 - 1})
\end{equation}
 so that we obtain
\begin{multline}
  \la W_1 W_2 W_3 \ra_{SU(2)}  = \frac{1}{Z} \frac{1}{8} \sum_{(w_1,w_2,w_3) = (\pm 1, \pm 1,\pm 1)} 
\sum_{n =1}^{\infty} 
\exp\left( - \frac{n^2}{8} a - \sum_{i=1}^{3} \frac{ (n + w_i)^2}{8} a_i
  \right) \frac{\prod_{i=1}^{3} (n + w_i)}{n} = \\
=\frac{1}{Z} \frac{1}{8} \sum_{(w_1,w_2,w_3)  \in S} \sum_{n \neq 0} 
 e^{ - \frac{n^2}{2} \alpha - n \beta - \gamma} ( n^2 + n c_1(w) +
 c_2(w) + \frac{1}{n} c_3(w))
\end{multline}
where
\begin{equation}
  \alpha = \frac{1}{4} (a + a_1 + a_2 + a_3) \quad
\beta = \frac{1}{4} \sum w_i a_i \quad 
\gamma = \frac 1 8  \sum w_i^2 a_i 
\end{equation}
and
\begin{equation}
  c_1(w) = \sum w_i \quad c_2 (w) = \sum_{i < j} w_i w_j \quad c_3(w) =
  w_1 w_2 w_3 
\end{equation}
and the index set $S = \{(1,1,1), (-1,1,1), (1,-1,1), (1,1,-1) \}$. 
Next we make a Poisson resummation (there is a subtlety as we have sum
over $n  \neq 0$ not over $n \in \BZ$ but the extra term $n=0$ actually
gives vanishing contribution after summation over S). We get
\begin{equation}
\label{eq:SU2-migdal}
  \la W_1 W_2 W_3 \ra_{SU(2)}  = \frac{1}{Z} \frac{1}{8}
  \sum_{(w_1,w_2,w_3)  \in S}
\sum_{m \in \BZ} (\p_\beta^2 - c_1(w) \p_\beta + c_2(w) - \p_\beta^{-1}
)  e^{\frac{1}{2\alpha} ( \beta + 2 \pi i m )^2 - \gamma }
\end{equation}
where $\p_{\beta}^{-1} F(\beta)  = \int_{0}^{\beta}  d \tilde \beta
F(\tilde \beta)$. The perturbative part is given by the term $m = 0$. 
We find 
 \begin{equation}
\label{eq:SU2-migdal-pert}
  \la W_1 W_2 W_3 \ra_{pert, SU(2)} =   1-\frac{3 \left(A (\sum A_i) + \sum_{i <j} 2 A_i A_j\right) g^2}{8 \left(A+A_1+A_2+A_3\right)} + \dots
 \end{equation}
If we wish to get the $U(2)$ correlation function, we simply need to multiply the above $SU(2)$
result by the $U(1)$ correlation function (\ref{eq:U1pert}) with
$g^2$ substituted  by $\frac 1 2 g^2$ (because in our conventions the Casimir of
$U(1)$ factor inside $U(N)$ is $\frac{1}{N}$ in fundamental representation). 

Starting from the matrix model (\ref{eq:MMaction}) for $U(2)$ (i.e. the matrix integral is over $2\times 2$ hermitean matrices), we have computed perturbatively $\langle W_F W_F W_F\rangle$ up to order $g^{10}$ and verified that it agrees with the expansion of the perturbative truncation (\ref{eq:SU2-migdal-pert}) (multiplied by the $U(1)$ contribution (\ref{eq:U1pert}) as described above) of the exact result (\ref{eq:SU2-migdal}). Of course, this agrees with the general proof given in section \ref{constant-F}. 

\section{Position independence of the $<WO>$ correlator from string theory}
Here we show analytically the independence of the integral in (\ref{bulk-int}) on the position $x_0$ of the local operator in the special case $\theta_0=\frac{\pi}{2}$ and ${\cal J}_1\rightarrow 0$. In this case the calculation reduces to $\langle W O\rangle $ and the non-trivial integral to compute is (see also \cite{Giombi:2009ds}) 
 \begin{equation}
I= \frac{1}{2 \pi} \int_{0}^{2 \pi} d \sigma \int_0^{\infty} \frac {d \tau}{\cosh^2 \tau}
\tanh^{j-2} \tau \left ( \frac {\cos \eta} { 1  - \sin \eta \cos \sigma \cosh^{-1} \tau}
\right)^j \stackrel{?}{=} \frac{1}{j - 1}
 \end{equation}
for any integer $j \geq 2$. 
The integral is elementary for $\eta = 0$ by the change of variables $\zeta =
\tanh \tau$. We want to show that, in fact, the integral does not depend
on $\eta$.
 In terms of the variables $ \sin \beta = \cosh^{-1} \tau, \cos \beta =
 \tanh \tau$ the integral
 is 
 \begin{equation}
2\pi  I=  \int_{0}^{2 \pi} d \sigma 
\int_{0}^{\frac \pi 2} d \beta \frac{ \sin \beta}{ \cos^2 \beta}
\left(
\frac{\cos \eta \cos \beta}
{1 - \cos \sigma \sin \eta \sin \beta}
\right)^j \,.
 \end{equation}
We now notice that the integrand
\begin{equation}
  f  = \frac{ \sin \beta}{ \cos^2 \beta}
\left(
\frac{\cos \eta \cos \beta}
{1 - \cos \sigma \sin \eta \sin \beta}
\right)^j 
\end{equation}
satisfies
\begin{equation}
 \cos \eta \, \p_{\eta} f\, d\sigma \wedge d \beta = d g
\end{equation}
where the one-form $g$ is
\begin{equation}
  g = f \cos \sigma \cos \beta \, d\sigma + f \frac{\sin \sigma} {\sin \beta} \, d \beta\,.
\end{equation}
Therefore 
\begin{equation}
  \cos \eta \, \p_{\eta} (2 \pi I) =  \int_{R} dg  = \int_{\p R} g = 0
\end{equation}
where the cylinder $R=I_{\beta} \times S^{1}_{\sigma}$ is the integration
domain 
and $\p R$ is the boundary consisting of two circles $S^{1}_{\sigma}$
 at $\beta =0$ and
$\beta = \frac{\pi}{2}$, and the last equality holds because $g$
identically vanishes for $\beta = 0$ or $\beta = \frac{\pi}{2}$. Hence
we have shown that $I$ does not depend on $\eta$ in the range
 $0 \leq \eta < \frac{\pi}{2}$ in which $f,g$ are smooth on
 the compact integration domain $R$. 
\bibliography{bsample}
\end{document}